\documentclass[manuscript]{aastex}

\usepackage{natbib}
\bibpunct{(}{)}{;}{a}{}{,} 

\usepackage{epsfig,times,lscape,rotating}
\usepackage{color}
\usepackage{amsmath}
\usepackage{amssymb}
\usepackage{txfonts}

\shorttitle{Global MHD Simulations of Stratified Protoplanetary Disks.}
\shortauthors{Flock et al.}

\begin{document}
\title{Turbulence and Steady Flows in 3D Global Stratified MHD Simulations of
Accretion Disks}

\author{M. Flock\altaffilmark{1}, N. Dzyurkevich\altaffilmark{1}, H. Klahr\altaffilmark{1},
N. J. Turner\altaffilmark{1,2},Th. Henning\altaffilmark{1}}
\affil{$^1$Max Planck Institute for Astronomy, K\"onigstuhl 17,
 69117 Heidelberg, Germany}
\affil{$^2$Jet Propulsion Laboratory, California Institute of Technology, Pasadena,
CA 91109, USA}

\begin{abstract}
We present full $2\pi$ global 3-D stratified MHD
simulations of accretion disks. We interpret our results 
in the context of proto-planetary disks.
We investigate the turbulence driven by the magneto-rotational instability (MRI) using
the PLUTO Godunov code in spherical coordinates with
the accurate and robust HLLD Riemann solver.  We follow the
turbulence for more than 1500~orbits at the innermost radius of the
domain to measure the overall strength of turbulent motions and the
detailed accretion flow pattern.\\  
We find that regions within two
scale heights of the midplane have a turbulent Mach number of about
0.1 and a magnetic pressure two to three orders of magnitude less
than the gas pressure, while outside three scale heights the
magnetic pressure equals or exceeds the gas pressure and the
turbulence is transonic, leading to large density fluctuations.
The strongest large-scale density disturbances are spiral density
waves, and the strongest of these waves has $m=5$. 
No clear meridional circulation appears in the calculations because
fluctuating radial pressure gradients lead to changes in the orbital
frequency, comparable in importance to the stress gradients that
drive the meridional flows in viscous models.  The net mass
flow rate is well-reproduced by a viscous model using the mean
stress distribution taken from the MHD calculation.
The strength of the mean turbulent magnetic field is inversely
proportional to the radius, so the fields are approximately
force-free on the largest scales.  Consequently the accretion stress
falls off as the inverse square of the radius.
\end{abstract}

\section{Introduction}
The magneto-rotational instability (MRI) is a powerful process to drive
turbulence and angular momentum transport in protoplanetary disks, 
ultimately enabling the accretion of matter onto the central object \citep{bal91,haw91,bal98}.
There is vast literature studying this mechanism in local shearing box
simulations with an ideal MHD description
\citep{bra95,haw95,haw96,mat95,sto96,san04}. \\
The effect of non-ideal MHD on MRI, regarding the issue of resistive 
protoplanetary disks, was mostly studied in
local box simulations \citep{bla94,jin96,san00,fle00,san02I,san02II,fle03,inu05,tur07,tur08,tur10}. 
The various studies showed that at a 
certain level of resistivity the MRI will be suppressed.

Up to now there is no prescription of resistive profile 
in protoplanetary disks which applies for longer timescales.
It is known that the dust grains control the ionization level
in protoplanetary disk. The particle cross section and dust-to-gas
ratio are most important parameters for defining the ionization level of
the gas \citep{tur06,war07}.
Most studies of non-ideal MRI turbulence use a static dust
distribution and neglect dust
growth and evolution \citep{sim09,tur06,tur07,dzy10}. But exactly in non-turbulent regions, the dust
 particles can grow, reducing quickly the cross section and so driving to
better ionization levels enabling again MRI \citep{zso10}. 
In our study we focus on ideal MHD, which applies to sufficiently
ionized disk regions depleted of small dust grains. 
This also applies to the innermost hot parts of 
protoplanetary accretion disk and even extended radial regions 
as expected for transitional disks \citep{chi07}.
Here, an MRI turbulent ionization front starting at the inner rim of the 
disk propagates radially outward and the disk get evacuated from
inside-out.\\

Recent results on the MRI obtained in local unstratified box simulations by \citet{les07},
\citet{froII07}, \citet{sim09} and \citet{fro10} show that occurrence and saturation level of MRI in zero net flux simulations is controlled by the magnetic Prandtl number \footnote{The ratio of viscosity vs.\
resistivity}. 
Ideal MHD simulations including stratification found at least convergence for zero-net flux local
simulations \citep{dav10,fla10}. Here, the vertical resolution plays the key role for the convergence
\citep{shi10}. Despite the simplicity of box simulations and their interesting results obtained over the
last years, the study of radial extended structures are very restricted.
For instance, the aspect ratio of the boxes is known to influence the saturation level of turbulence \citep{bod08}.
Besides local box simulations, global simulations of MRI have also been performed
\citep{arm98,haw01I,ste01,haw01,arl01,fro06,lyr08,fro09}.
They confirmed the picture of a viscously spreading disk as a proxy for the action of MHD turbulence. 
Recently, the first global non-ideal MHD simulation \citep{dzy10}, 
which included a radial dead-zone / active zone interface demonstrated the importance of the inner edge of the
dead zone as a trap for planetesimals and even small planets.\\

So far only finite difference schemes as implemented in the ZEUS and Pencil
codes have been used to perform global simulations. 
However, a Godunov code would have several advantages over a finite difference scheme. 
Without using any artificial viscosity the code solves 
the MHD Riemann problem and can better handle the supersonic MHD turbulence in corona regions of the disk. Several papers recognized the importance of Godunov-type shock capturing upwind schemes for future astrophysical 
simulations \citep{sto05,fro06b,mig07,flo10}.
In this project we use the Godunov code PLUTO \citep{mig09} to perform
isothermal global MHD simulations of protoplanetary disks. 
In future work, we
will switch on the total energy conservation property of this scheme and
include radiation MHD to follow the temperature evolution in global disk.\\
An important issue in stratified global simulations is the
relatively low resolution per scale height compared to what is possible in local box simulations.
In addition, the extent of the azimuthal domain is often restricted to save computational time.
The first $2\pi$ global disk simulations were performed by
\citet{arm98}, \citet{haw00} and
\citet{arl01} for a short period of time.
But most of the global simulations were performed in restricted azimuthal domains like
$\pi/4$ or $\pi/2$ \citep{fro06,fro09,dzy10}. 
By restricting the azimuthal domain one also restricts the largest possible mode in the
domain. One of the goal of this paper is to investigate whether these
modes affects the nonlinear state of the turbulence.\\

The standard viscous $\alpha-$disk theory \citep{sha73} introduces an effective turbulent viscosity 
 $\eta = \rho \nu = \alpha P / \Omega$ with the local thermal pressure $P$
and the orbital frequency $\Omega$,
arising from undefined magnetic or hydrodynamic processes,
transporting angular momentum outward and allowing mass accretion onto the star.
\citet{lyn74} calculated the radial mass accretion rate and the
radial accretion velocity for a 1D viscous disk model as a local function of surface density $\Sigma$ and the value for $\alpha$.
Interestingly 2D viscous disk models \citep{kle92} showed the appearance of meridional
outflows. Here, the mass flows radially outward near the
midplane compensated by increased radial inflow at upper layers of the disk to allow for net-accretion,
for $\alpha < 0.05$. Much emphasis was given to this radial outflow and its role
for the radial transport for grains and chemical species over large distances and relative short time scales
\citep{kel04,cie09}.\\
In addition, we will investigate the onset of a vertical outflow
as it was described in local box simulations \citep{suz09,suz10} using a net
flux vertical field. Such outflows can be launched in the
magnetized corona region of the MRI turbulent disk \citep{mil00,mac00}.
They could have an important effect on the dissipation timescales of
accretion disks and may be related to jet production from accretion
disks \citep{fer06}.
An interesting property of MRI in stratified zero-net flux simulations is the emergence of
a "butterfly" pattern, an oscillating mean azimuthal magnetic field with an
period of 10 local orbits. It was found in many local MRI simulations,
recently again by \citet{dav10}, \citet{gre10}, \citet{fla10} and in global
simulations by \citet{sor10} and \citet{dzy10}.
We indeed identify such a "butterfly" pattern in our global runs, which was suggested to be linked to magnetic dynamo action in accretion disks \citep{sor10,gre10}.

Our paper is organized in the following way.
In Section 2 we will present our model setup and the numerical configuration.
Section 3 contains the results 
Section 4 and 5 will provide a discussion, summary and outlook for our work.
\section{Model setup}
The setup follows closely the disk model which is presented by
\citet{fro06,fro09}.
We define the cylindrical radius with $R = r \sin{(\theta)}$ with the spherical
radius $r$ and polar angle $\theta$.
The initial density, pressure and azimuthal velocity are set to be in hydrostatic equilibrium.
$$\rho = \rho_{0}  R^{-3/2}\exp{}\Bigg(\frac{\sin{(\theta)}-1}{(H/R)^2}\Bigg) $$
with $\rho_{0} = 1.0$, $\rm H/R = c_0 = 0.07$.\\ 
We choose an isothermal equation
of state. The pressure is set to $P = c_{s}^2\rho$ with $\rm c_{s} =
c_0\cdot1/\sqrt{R}$.
The azimuthal velocity follows  $$V_{\phi} = \sqrt{\frac{1}{r}}\Bigg(1- \frac{2.5}{\sin(\theta)}c^2_0 \Bigg).$$
For the initial velocities $V_{R}$ and $V_{\theta}$ we use a white noise
perturbation amplitude of $V_{R,\theta}^{Init} = 10^{-4} c_{s}$.�
We start the simulation with a pure toroidal magnetic seed field with constant plasma beta
$\beta = 2P / B^{2} = 25$.
%
The radial domain extends from 1 to 10 radial code units (CU)\footnote{We refer to CU instead of a physical length unit because ideal MHD simulations without radiation transport
are scale free. Thus our simulations could represent a disk from 1 to 10 AU as much as a disk from
$0.1$ to 1 AU. Only explicit dust physics and radiative transfer will introduce a realistic physical scale.} with radial buffer zones from 1 to 2 CU
and 9 to 10 CU.
In the buffer zones we use a linearly increasing resistivity. This damps 
the magnetic field fluctuations and suppresses boundary interactions,
especially for the closed boundary runs. Our buffer
zone follows mainly the ones used in global simulations by \citep{fro06,fro09,dzy10}. 
The $\theta$ domain is set to $\theta = \pi/2 \pm 0.3 $, corresponding to $\pm
4.3 \rm $ scale heights.
We calculated in total five disk models. Three models cover the complete $2\pi$ azimuthal  domain (FC, FO and BO in Table 1) and 
two models are constrained to $\pi/4$ (PC, PO in Table 1). The simulation FO
is also used for a test model FOR which is described later.
The simulation BO has the best resolution.
One subset of models has a closed boundary (FC and PC). Here we
use a reflective radial boundary with a sign flip
for the tangential magnetic fields and a periodic boundary condition for the $\theta$ direction.
A second subset of models has an outflow boundary condition (FO,PO and BO). Here, we use a relaxation function in the radial buffer zones which reestablishes gently the initial value of density over a time period of one local orbit. 
In the buffer zones we set: $\rho^{new} = \rho - (\rho-\rho^{Init})\cdot \Delta
t / T_{Orbits}$. 
Our outflow boundary condition projects the radial gradients
in density, pressure and azimuthal velocity into the radial boundary and the
vertical gradients in density and pressure at the $\theta$ boundary. 
We ensure to have no inflow velocities. For an inward pointing velocity
we mirror the values in the ghost cell to ensure no inward mass flux. 
The $\theta$ boundary condition for the magnetic field are also set up 
to be zero gradient, which approximates "force-free" - outflow conditions. 
We also ensure the force free character of the tangential components for the radial boundary
by adjusting the $1/r$ profile in the magnetic field components in the ghost
cells.
The normal component of the magnetic field in the ghost cells is always 
set to have $\nabla \cdot \vec{B}$ = 0.
We set the CFL value to 0.33. Also higher CFL values were successfully tested and
will be used for future calculations.
We use a uniform grid with an aspect ratio of the individual cells at 5 CU of $1:0.67:1.74$
$(\Delta r: r\Delta\theta:r\Delta\phi)$.
Using a uniform grid instead of a logarithmic grid, where
$\Delta r/r$ is constant, has the disadvantage that it will reduce the
accuracy in the sense that the inner part of the disk is poorly resolved,
compared to the outer part of the disk: $H(1AU)/\Delta r < H(10AU)/\Delta r  $.\\ 
However for the uniform grid, the relative broad radial inner 
buffer zone lies in the poorly resolved disk part and is excluded from
analysis.
The outer parts of the disk are, compared
to a logarithmic grid with the same resolution, better resolved.  
Logarithmic grid requires a much smaller buffer
zone, e.g. a logarithmic grid would place one third of the total number of grid
cells in the first ninth of the domain, between 1 and 2 AU.
Of course, using a uniform grid will always restrict the range of the
radial domain and for more radially extended simulation the need of a logarithmic grid is mandatory.

For all runs we employ the second order scheme in
PLUTO with the HLLD Riemann solver \citep{miy05}, piece-wise linear
reconstruction and $2^{nd}$ order Runge Kutta time integration. 
We treat the induction equation with the "Constrained Transport" (CT) method in combination with the upwind CT method described in
\citet{gar05}.
The detailed numerical configuration is presented in \citet{flo10}.
Our high resolution run BO was performed on a Blue-gene/P cluster with 4096
cores and was calculated
for over 1.5 million time steps which corresponds to 1.8 million CPU hours.
\begin{table}[th]
\begin{center}
\begin{tabular}{ccccc}
Model name & Resolution ($R$ $\theta$ $\phi$)  & $\phi$-range & Boundary & Orbits at 1 AU (Years)\\
\hline
\hline
PC & 256 128 64 & $\pi/4$ & closed & 1435 \\
PO & 256 128 64 & $\pi/4$ & open & 1519 \\
FC & 256 128 512 & $2\pi$ & closed & 1472 \\
FO & 256 128 512 &  $2\pi$ & open & 1526 \\
\hline
FOR& 256 128 512 & $2\pi$ & open & $1000-1448$ \\ 
\hline
BO& 384 192 768 & $2\pi$ & open & $1247$ 
\\
\end{tabular}
\caption{MHD runs performed. (P - $\pi/4$; F - $2\pi$; O - open boundary; C -
closed boundary; B - best resolved run.)}
\end{center}
\end{table}
\subsection{Code Units vs. Physical Units}
Isothermal ideal MHD simulations are scale-invariant. 
One has to define unit-variables to transform from code to cgs units. 
We can set three independent values to 
define our problem. 
This is gas density for which we choose for instance $\rm \rho_u = 10^{-10} g/cm^3$, the radial
distance unity as length $1 CU = 1 AU$
and the Keplerian velocity $\rm v_u = \sqrt{ G\cdot M_{\sun} / l_u}$ with the gravitational
constant $G$ and the solar mass $M_{\sun}$.

With those three quantities, we translate the values for our measured surface density and mass accretion
rate into cgs units. Using this values we derive at 1 AU at the midplane a gas density of
$\rm \rho = 10^{-10} g/cm^3$ with a Keplerian velocity of $\rm v_K= 2.98* 10^6 cm/s$. 
With this the surface density becomes $524 g/cm^2$ at 1AU.
Gas velocities and the Alfv\'enic speed are always presented in units of the sound
speed for convenience.

\subsection{Turbulent stresses}
The $\alpha$ parameter relates the turbulent stresses to 
the local thermal pressure. For the calculation of the $\alpha$ values
we measure the Reynolds and Maxwell
stresses, which are the $R-\phi$ components of the respective stress tensors. 
The Reynolds stress is calculated as $\rm T_{R}= \overline{\rho v'_{\phi}v'_{R}} $
and the Maxwell stress as $\rm T_{M}= \overline{B'_{\phi}B'_{R}}/ 4 \pi $
with the turbulent velocity or magnetic fields, e.g., $\rm v'_{\phi} = v_{\phi} - \overline{v_{\phi}}$.
The mean component of the velocity and magnetic field are always calculated
only along the azimuthal direction because of the radial and vertical gradients in the disk.
In our simulations, the amplitude of Maxwell stress is about three times the Reynolds stress.
For the total $\alpha$ value we integrate the mass weighted stresses over the total domain
$$ \alpha = \frac{ \int \rho \Bigg( \frac{v'_{\phi}v'_{R}}{c^2_s} - \frac{B'_{\phi}B'_{R}}{4 \pi \rho c^2_s}\Bigg)dV} {\int \rho dV}.$$
The respective turbulence enhanced viscosity can now be represented as $\nu = \alpha H c_s$
with the height of the disk $H$ and the sound speed $c_s$.
\begin{figure}
\hspace{-0.6cm}
\begin{minipage}{5cm}
\psfig{figure=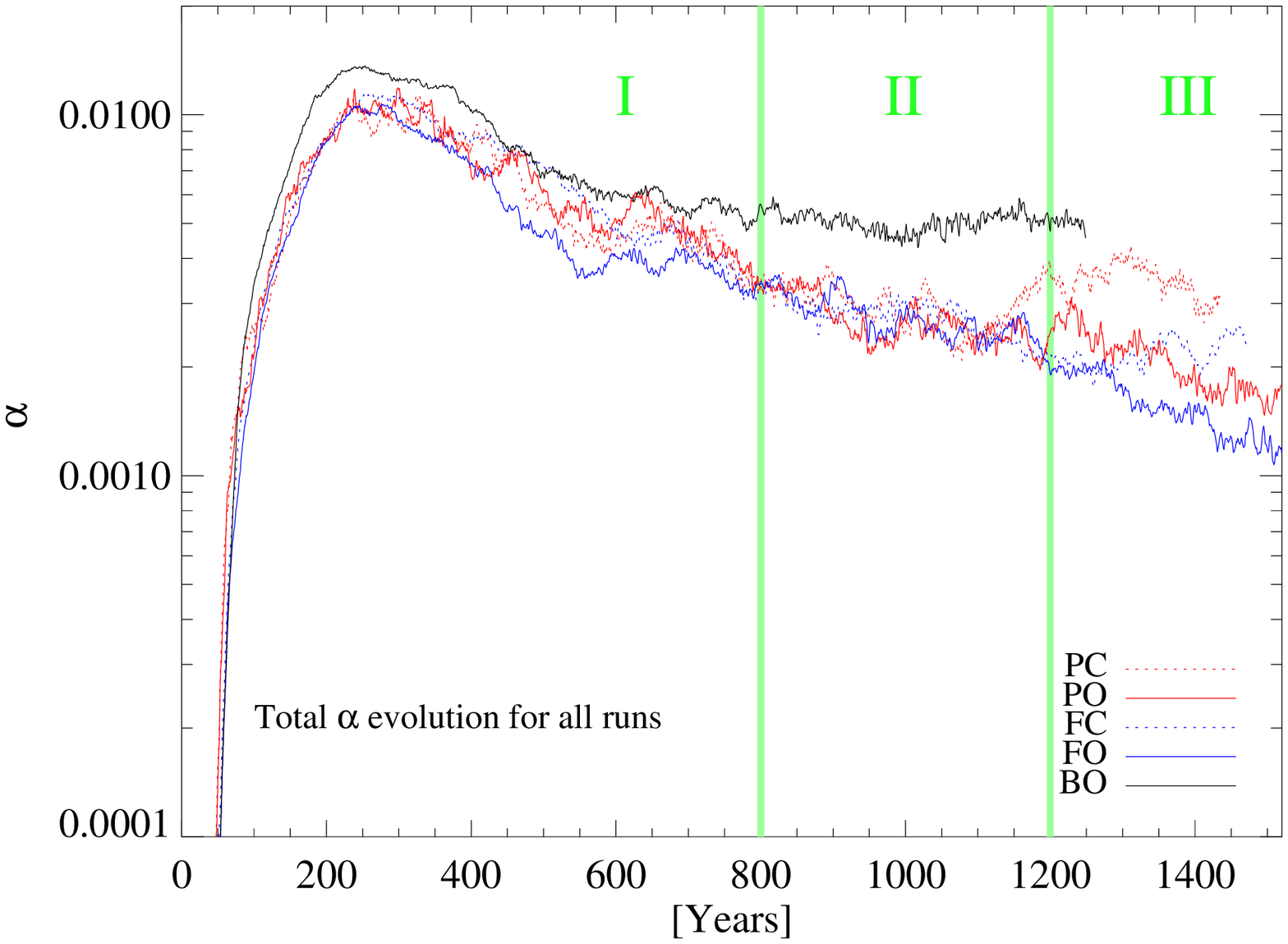,scale=0.46}
\psfig{figure=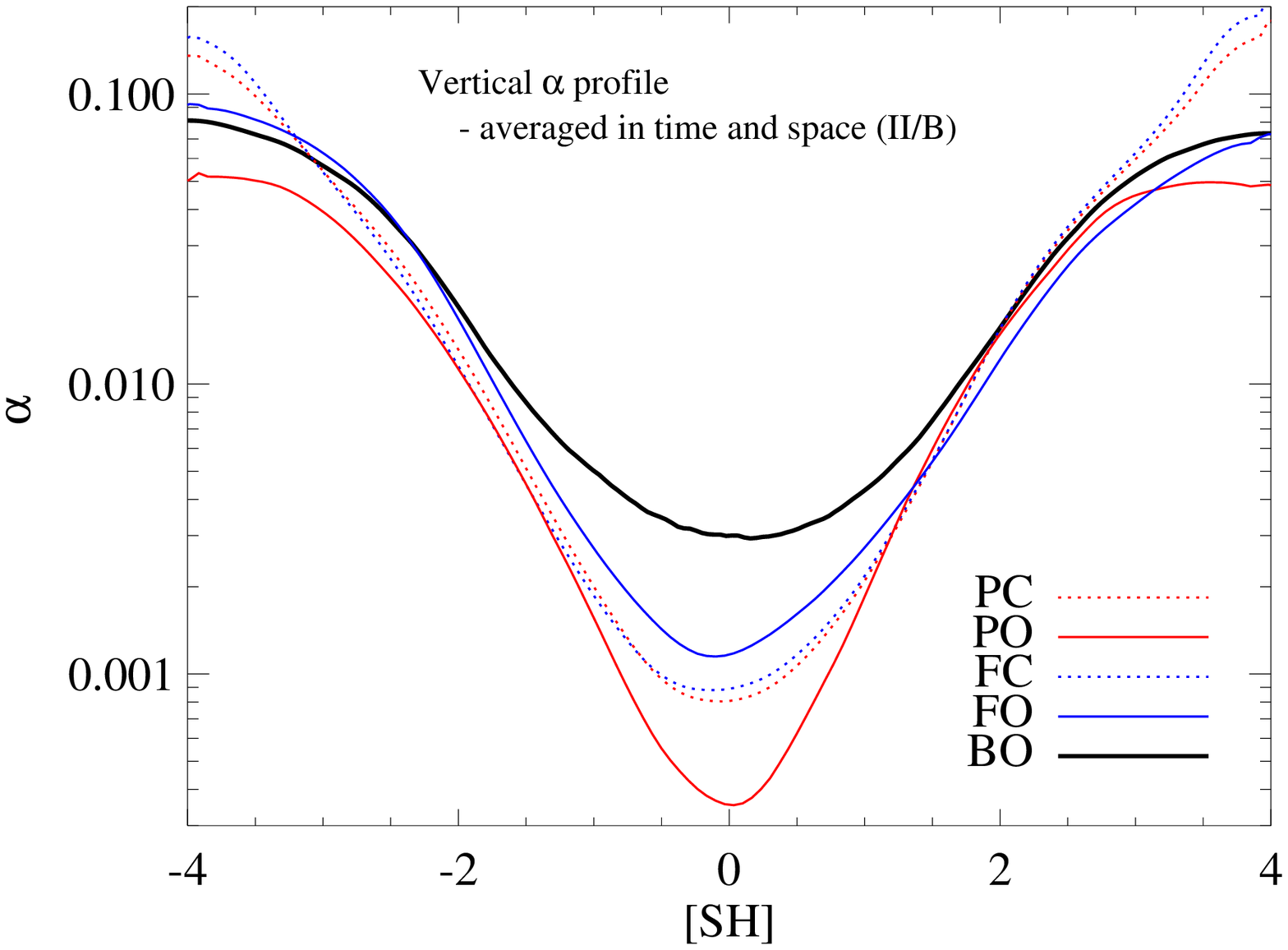,scale=0.46}
\end{minipage}
\hspace{4.0cm}
\begin{minipage}{5cm}
\psfig{figure=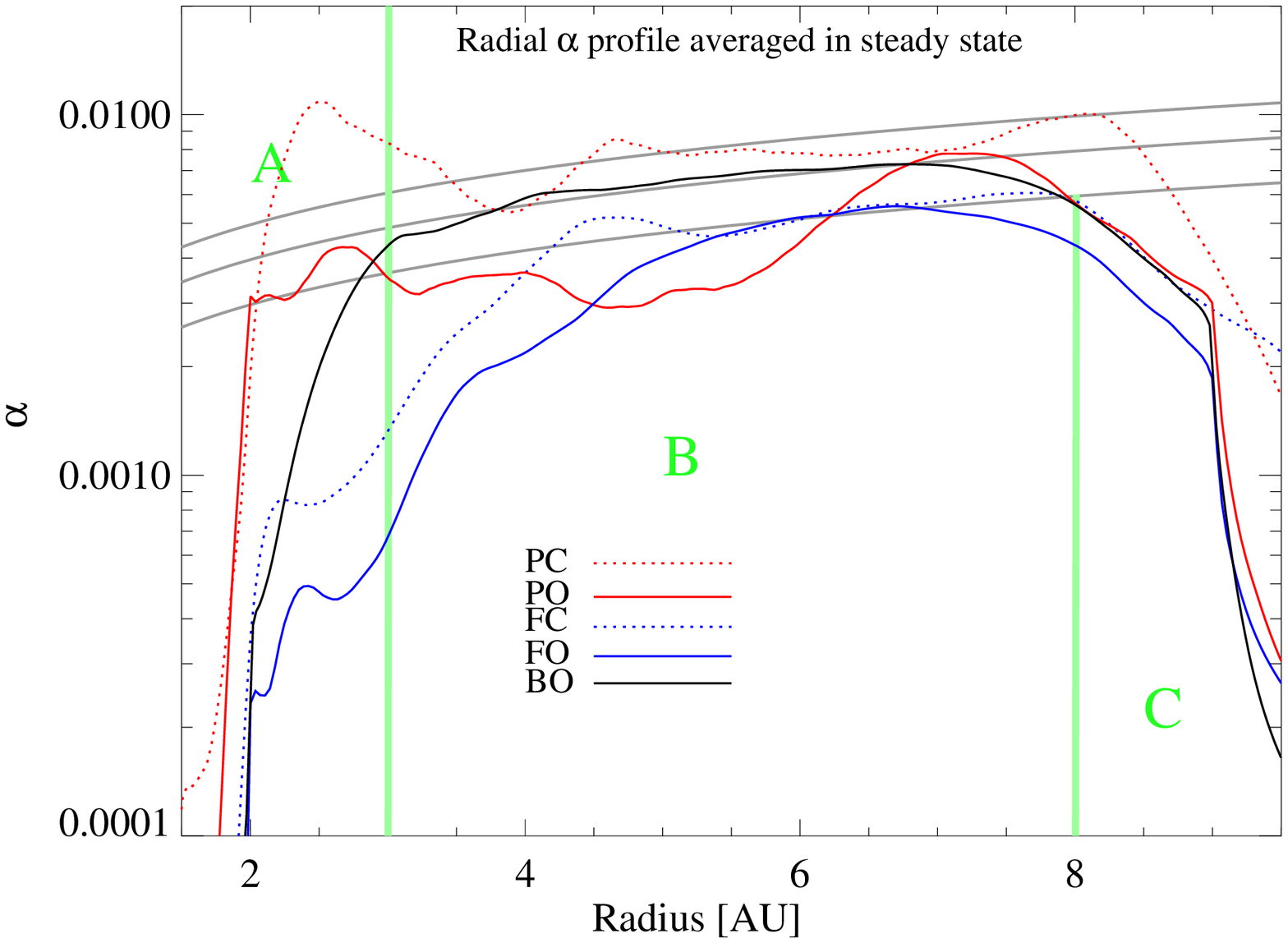,scale=0.46}
\psfig{figure=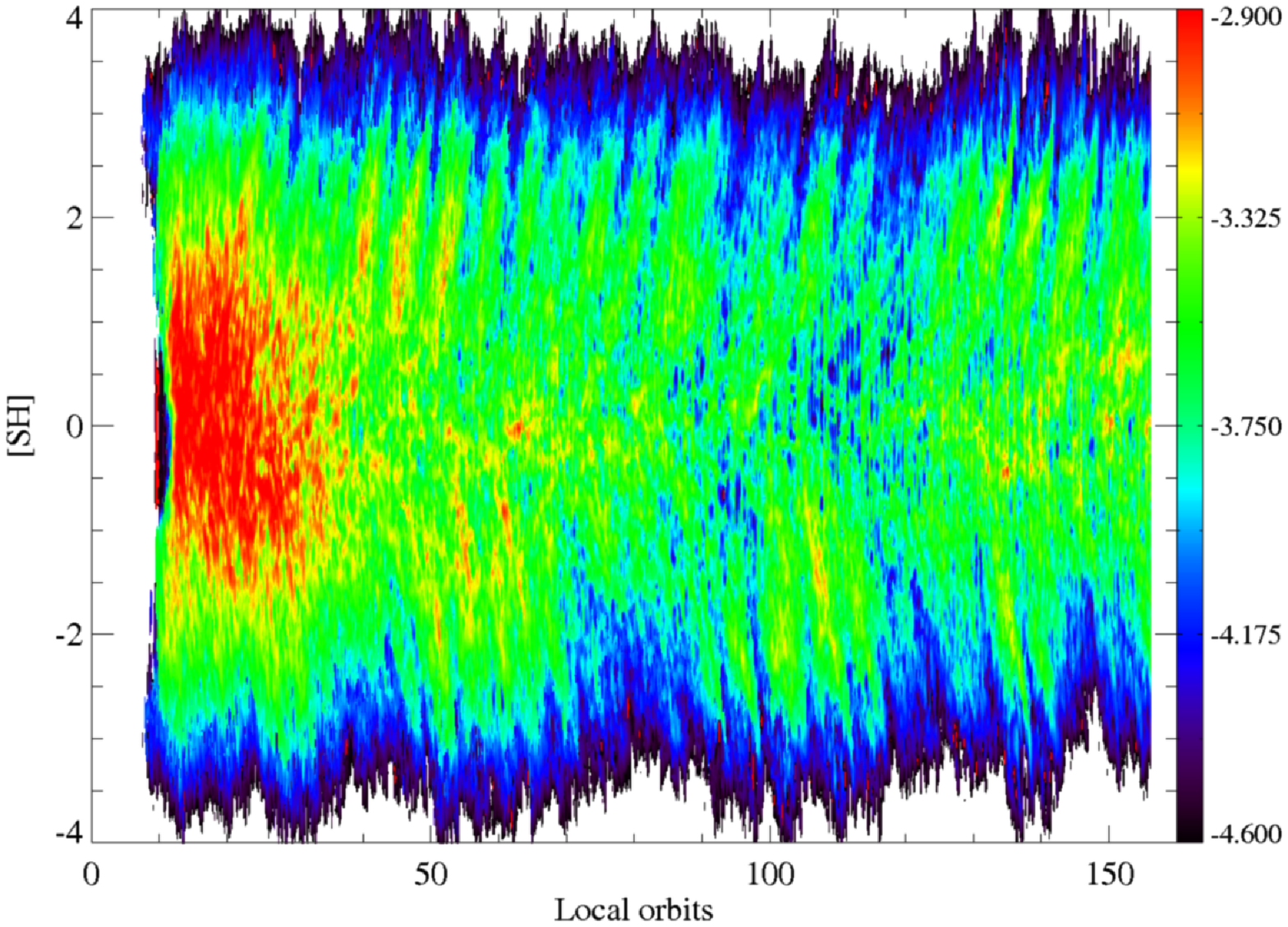,scale=0.46}
\end{minipage}
\label{totalal}
\caption{Top left: Total stresses expressed as $\alpha$-parameters for the $\pi/4$ models PC
and PO and the $2\pi$ models FC, FO and BO. The parameter is mass weighted
and integrated for the domain (3-8 AU).
Top right: Radial $\alpha$ profile, time averaged between 300 and 1200 inner
orbits.
For the best resolution model BO, the profile follows roughly $\sqrt{r}$ in the region B. 
Region A and C are affected by boundary conditions and buffer zones.
Bottom left:  Vertical $\alpha$ profile, averaged over time and space region ($II/B$). 
Bottom right: Evolution of the vertical distribution of azimuthally averaged Maxwell and Reynolds stress
$T_S$ for a radius of 4 AU with
time. Colors are logarithmic values of the corresponding dynamical viscosity including the density profile.}
\end{figure}
\section{Results}
\subsection{Disk evolution}
We first describe the typical evolution for an azimuthal MRI (AMRI) 
in global disk simulation with open boundaries. 
The AMRI simulation starts with a purely toroidal net magnetic field which 
becomes MRI unstable on timescales of around 10 local orbits.
After approximately 250 inner orbits, the disk reaches its maximum
$\alpha$ value of $0.01$.
At this time (equivalent to 10 local orbits at
the outer boundary of the undamped region) the disk has become fully turbulent.
During this evolution, the initial magnetic flux decreases as it leaves the computational
domain in the vertical direction. 
Starting at approximately 250 inner orbits, there is an evolution of oscillating mean fields.
Fig. 1, top left, presents the mass weighted and domain integrated $\alpha$ value 
over time for all models.
During the time period between 800 and 1200 inner orbits, we get a
relatively constant $\alpha$ value of $5\cdot10^{-3}$ (model BO). 
We mark three different time stages of the turbulent
disk evolution:
In period I (0 to 800 inner orbits), the turbulence is not yet saturated.
After a strong initial rise due to the net azimuthal field the turbulent decays
to a level where self-sustained turbulence is possible, e.g., the loss of magnetic flux
in the vertical direction is balanced by the generation of magnetic flux in the turbulent flow.
The nature of this generation of magnetic fields can be an indication for dynamo action,
e.g., an $\alpha \Omega$ dynamo
\citep{bra05,gre10}, but detailed studies of this effect will be subject to future work.
In period II, we have a quasi steady state of at least 400 inner orbits. 
During this time period, we do all our analysis. 
In the period III, a comparison of the models becomes less useful.
The models with lower resolution and open boundary (PO and FO) 
show a decreasing $\alpha$-stress in time. Thus they are not useful for long time integrations past
1200 inner orbits. In the closed models, on the other hand, the magnetic flux cannot escape vertically 
(PC and FC) and therefore, turbulence does not decay. 
On the contrary, turbulence even increases in these runs as the flux in the box 
cannot efficiently escape.\\
A closer view shows that the stress can oscillate locally on shorter time  scales.
In Fig. 1, bottom right, we plot the mass weighted stresses $\alpha\rho$ at 4 AU over height
and local orbits. The strength of stresses locally oscillates with a period of around 5 local
orbits. The maxima in the stresses always appear first in the midplane and then propagate
vertically. These oscillations in the stresses are connected to the $B_\phi$ "butterfly" structures
where the azimuthal mean field oscillates with a frequency of 10 local orbits (Fig. 13). 
Every change of sign in the mean $B_\phi$ is now correlated with a minimum in the stresses, 
which both occur every 5 local orbits.
At the same time this plot shows the importance of the stresses in a region up
to 3 disk scale heights.

Although one can always define a total $\alpha$-parameter in the disk, the spatial
variations are enormous, especially in the vertical direction. 
The vertical $\alpha$ profile is plotted in Fig.1, bottom left. For model
BO, the turbulent stress at the midplane increases from $2.0\cdot 10^{-3}$
up to $8\cdot10^{-2}$ at 4 scale heights.
The simulations with moderate resolution show significant lower values around the midplane due
to the lack of resolution there. For the closed models, the stresses in the
corona are artificially increased due to the periodic boundary.\\

\subsection{Radial profile of turbulent stress}
Beside the vertical profile of turbulent stress, which has been already studied in local
box simulations, the radial profile of the turbulent stress can only be addressed in global
simulations.
In Fig. 1, top right, we present the radial $\alpha$ profile, averaged
between 300 and 1200 inner orbits.
In the inner buffer zones (1 - 2 AU) the $\alpha$ values are practically zero because of the
resistive damping. Starting from 2 AU $\alpha$ rises until it levels off at
around 3 AU. From 3 to 8 AU we obtain a radial $\alpha$ profile which can be
approximated by a $\sqrt{r}$ dependence. Beyond 9 AU $\alpha$ is again close to zero because of the damping applied there.
We mark three regions in radius (Fig. 1, top right, green lines).
Region A, extending from 1 to 3 AU, is affected by the buffer zone.
Region B, ranging from 3 to 8 AU shows the $\sqrt{r}$ slope.
Region C, covering 8 to 10 AU, is again affected by the buffer zone.
In the following analysis, we will therefore concentrate on region B.\\
In order to have a radial force-free accretion disk, 
fields have to drop radially as $B \propto r^{-1}$ (Fig. 12, top left).
This was also observed for magnetic fields in galactic disks \citep{bec01} (Fig. 1).
If the most important toroidal field follows $\propto r^{-1}$, 
the radial Lorentz force vanishes:
$$\rm F_{radial} = - \frac{1}{r^2\rho} \frac{\partial r^2 B_{\phi}^2}{\partial
r}.$$
In the case of $\rm \partial \log{\rho}/\partial \log{r} = -1.5$
and $\partial \log{c_s}/\partial \log{r} = -0.5$ the $\alpha$ value,
dominated by the Maxwell stresses will then 
scale as $\sqrt{r}$, which is actually matching the value that 
we measure in our best resolved model (see Fig. 1, top right).

\subsection{Mass loss}
The models with open boundaries show a considerable mass loss over the course of our
simulations.
A vertical outflow removes a substantial amount of mass. The
total mass loss over time is presented in Fig. 5. The mass loss is
determined in space region B. The closed models FC and PC loose there mass
due to radial mass movement. The open models loose there mass mainly due to
the vertical outflow. We will discuss this outflow in section 3.5 .
To check the possible impact of this mass loss onto the properties of the
turbulence, we restarted run FO after 1000 inner orbits with the current
velocity and magnetic field configuration but the initial density distribution.
We call this model FOR (FO Restarted, Table 1).
After restarting the simulation, the turbulence needs a couple of
inner orbits to readjust the fully turbulent state.
We compared the mean total $\alpha$ stress of the runs FO and FOR and found a comparable evolution 
of the $\alpha$ values.  We measure $\alpha = 1.4\cdot10^{-3}$ for FOR and $\alpha = 1.3\cdot10^{-3}$ for FO in the time period
from 1000 up to 1400 inner orbits. We conclude that the mass loss is not
yet influencing the development and strength of turbulence. 
\begin{figure}
\hspace{-0.6cm}
\begin{minipage}{5cm}
\psfig{figure=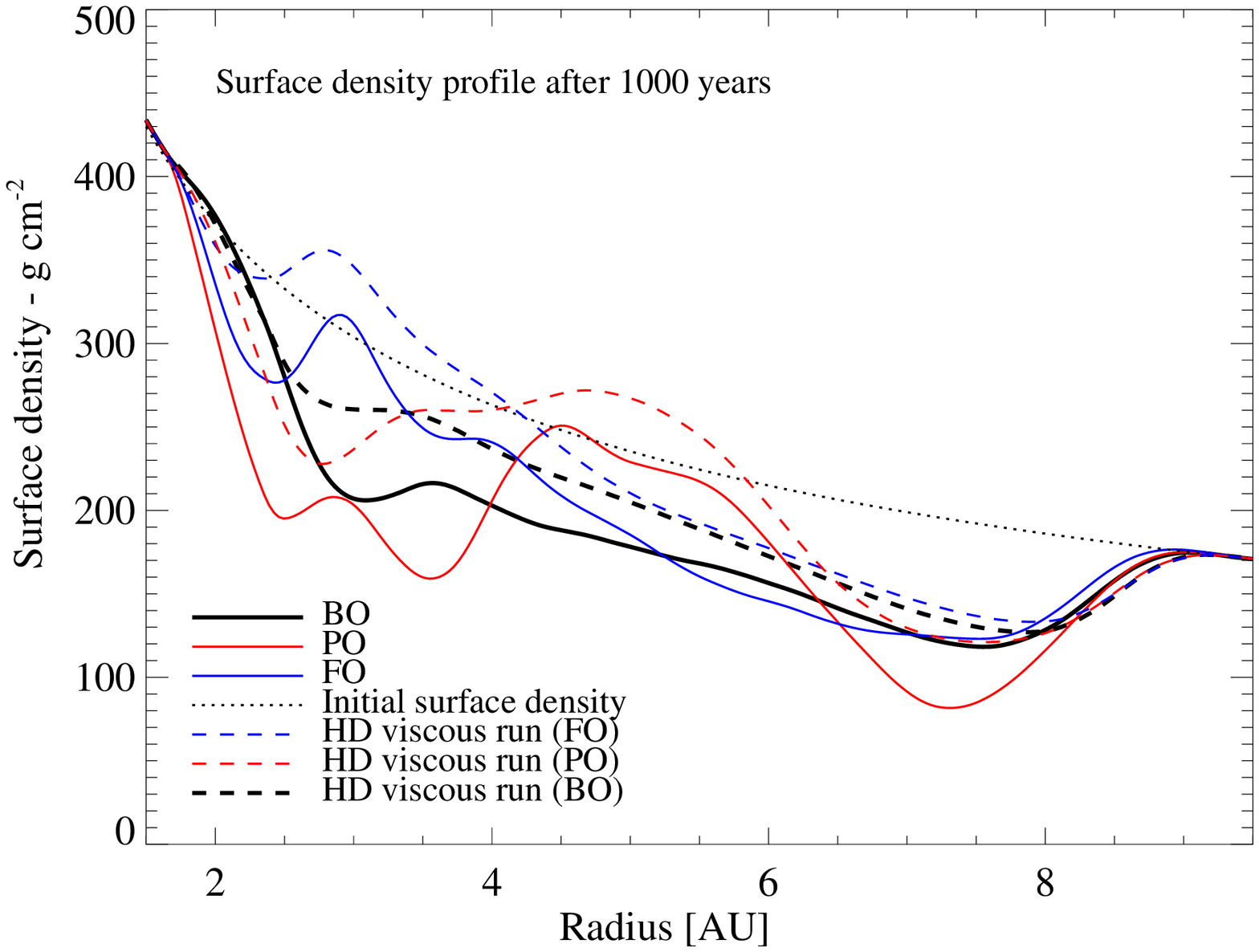,scale=0.46}
\psfig{figure=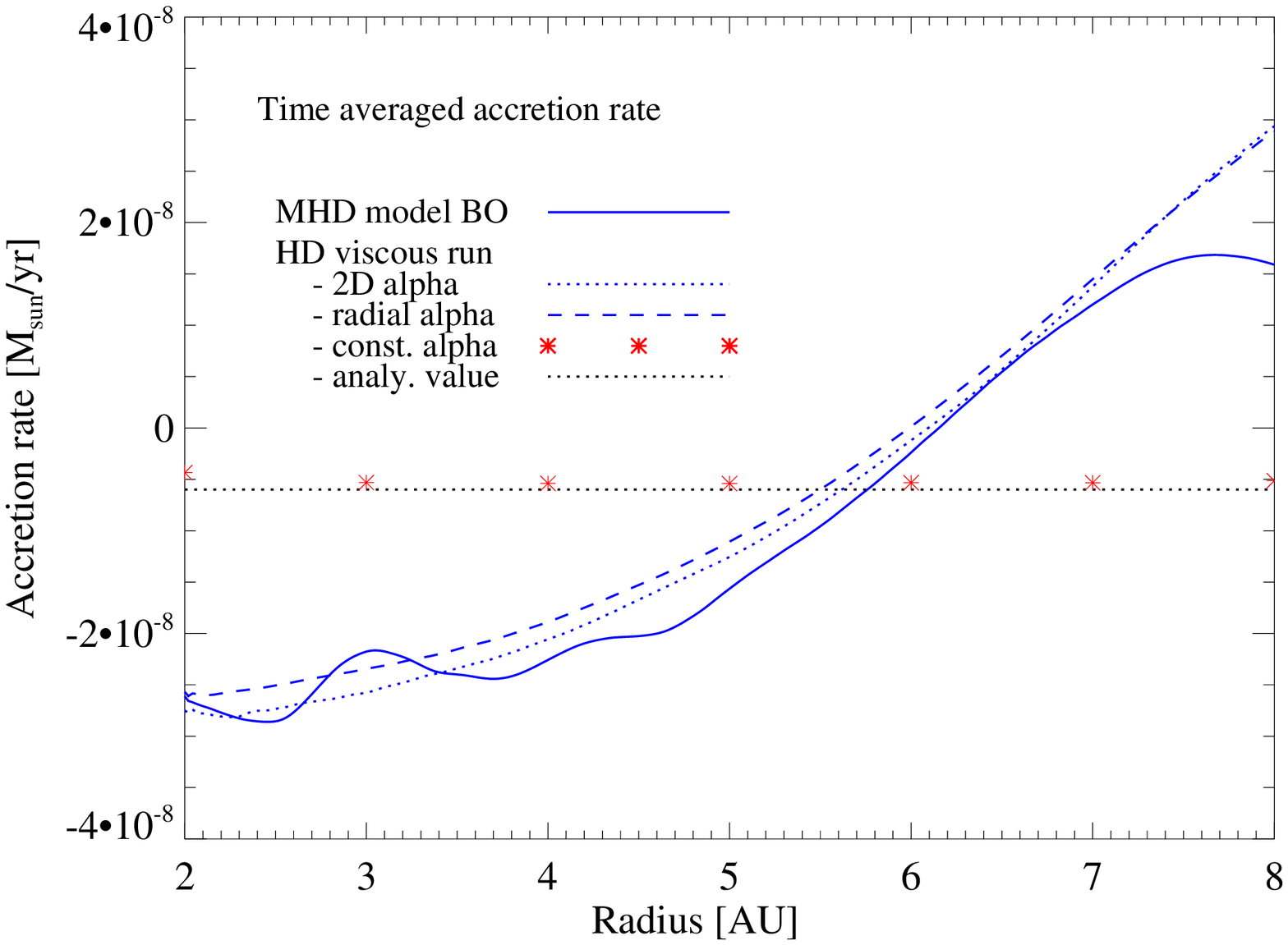,scale=0.46}
\end{minipage}
\hspace{4.0cm}
\begin{minipage}{5cm}
\psfig{figure=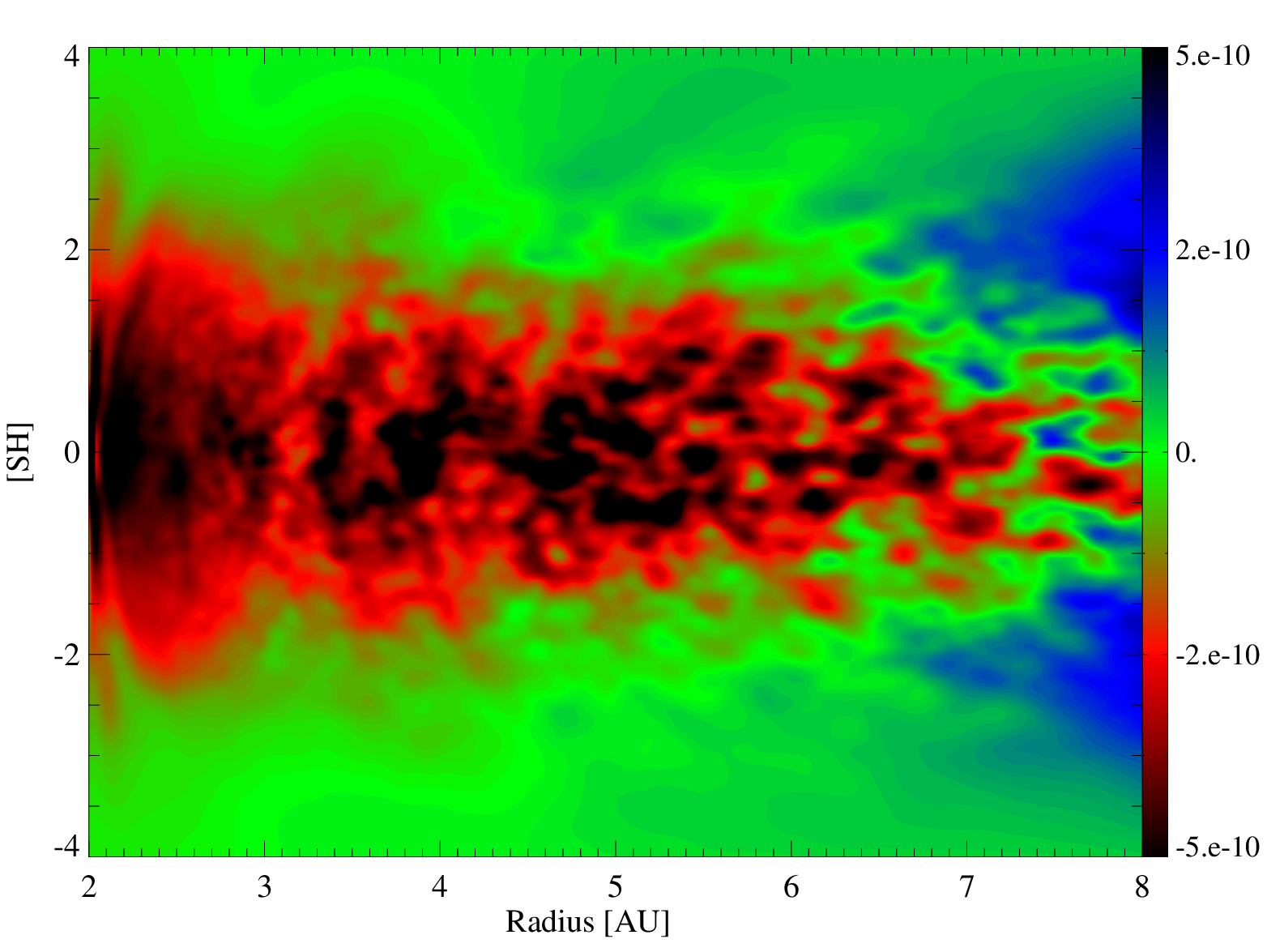,scale=0.46}
\psfig{figure=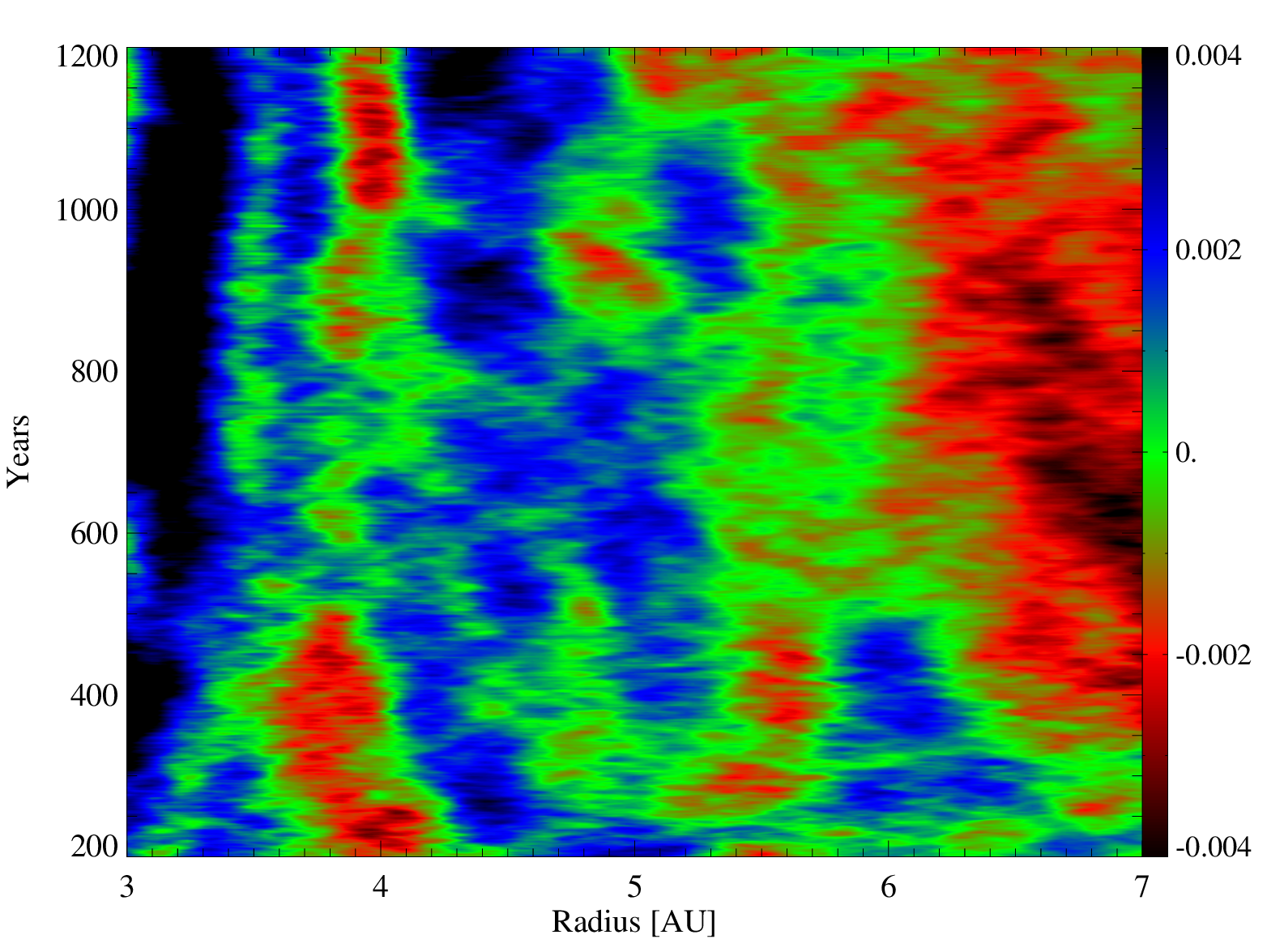,scale=0.46}
\end{minipage}
\label{surf}
\caption{Top left: The surface density profile after 1000 inner orbits for the $\pi/4$ model
PO, the $2\pi$ model FO and the high-resolution model BO. Dashed lines represent the surface
density profile for the respective viscous disk model and the dotted line the initial profile. 
Top right: 2D contour plot of time and azimuthal averaged radial mass flow for model BO.
The red color indicates inward accretion to the star, blue color shows outward motion.
We do not observe a meridional flow.
Bottom left: Radial profile of the time-averaged radial mass flow for
the high-resolution model BO (solid line) and the viscous models. 
Bottom right: 2D contour plot of $(\Omega - \Omega_0) / \Omega_0$ over
radius and time, averaged over azimuth at the midplane. The orbital frequency
remains sub-Keplerian $(\Omega_K - \Omega_0) / \Omega_0 = 0.012$.}

\end{figure}
\subsection{Viscous disk models}
The classical $\alpha$ viscous disk model should reproduce the radial mass flow 
as it occurs in global MHD simulations of MRI turbulent disks.
\citet{bal99} have
argued that the mean flow dynamics in MRI turbulence follows the
$\alpha$ prescription. 
To further test this supposition, we performed a series of 2D HD viscous comparison simulations with the PLUTO code 
for several of our 3D MHD runs. 
We use the same resolution and the same initial setup. 
Yet, the magnetic field evolution is now replaced by an explicit shear 
viscosity from the time averaged radial $\alpha$ profile $\nu (R) = \alpha(R) H c_s$
obtained from the MHD simulations (see Fig. 1 top right).
Fig. 2, top left, shows the surface density profile for the models PO,
FO, model BO and the corresponding viscous
model.
The surface density profile of the viscous runs follow nicely the respective MHD model
profile (Fig. 2, top left, dashed line) for the region that we use for analysis (3-8 AU).
All viscous models show a higher surface
density profile than the open models BO, PO and FO, but of course
in contrast to the MHD models the viscous models do not show any 
substantial vertical mass outflow.
The total radial mass flow (e.g. azimuthally and vertically integrated) is plotted as
a time average (0 - 1000 inner orbits) for 
the high-resolution model B0 (Fig. 2, bottom left, solid line) and the respective viscous
runs (Fig. 2, bottom left, dashed and dotted line). 
The radial mass flow of the viscous run matches very well the flow obtained in the MHD model.
A constant $\alpha$ value will not reproduce the proper
evolution of the MRI run. If we adopt for instance a constant $\alpha$ value of $5\cdot10^{-3}$, 
which would be the global mean value of the MHD run, we get a globally constant accretion rate of
$\rm 5.1\cdot10^{-9} M_\sun/yr$.
As a sanity check for our viscosity module we compare this value to the 
analytical estimates by \citet{lyn74}: 
$$\dot M(r) = 3 \pi \Sigma_g \nu + 6 \pi r  \frac{\partial(\Sigma_g\nu)}{\partial r}$$
and find a value very close to the time dependent viscous run of $\dot M = 6\cdot10^{-9}
\frac{M_{\sun}}{yr}$,
based on a surface density profile of $\rm \Sigma_g = 524\cdot R^{-0.5} [g/cm^2]$ and our disk parameter
$H=0.07*R$.\\
In Fig. 2, top right, we show the time and azimuthal average of 
the accretion rate over radius and height. 
There is a dominant inward accretion at the midplane (Fig. 2, top right, red
color).
This result is in contrast to the viscous runs where we see the minimum of
accretion and even a small outflow at the midplane \citep{kle92,tak02}.
After \citet{tak02} (eq. 8) there are several possibilities which could change
the vertical profile of the radial velocity, and therefore the mean accretion
flow. Radial and vertical gradients in the orbital frequency as well as a
spatially varying $\alpha$ will affect the vertical profile of the meridional outflow.
For the MHD simulations one has to include the vertical gradient as well as the time derivative of the orbital
frequency. Fig. 2, bottom right, demonstrate the change of the orbital frequency
with a period of around 50 local orbits at 5 AU.\\
Radial mass flow and surface density evolution have shown that we can fit
our MHD global models with a viscous disk model as long as we use 
an $\alpha$ profile compatible with our MHD run.
Of course, the disk spreading that we observe in our MHD run
is partly due to the existence of our radial buffer zones,
in which not only the fields decay, but also the $\alpha$ stresses
vanish. In a larger radial domain we can expect that also a larger
region of the disk will get into a steady state of accretion.
However, one could also argue that in a realistic protoplanetary accretion disk
one will ultimately reach dead zones which behave similar as our
buffer zones. In that sense the active part of our global disk is 
embedded between two dead zones. 
%
\begin{figure}
\hspace{-1.2cm}
\begin{minipage}{5cm}
\psfig{figure=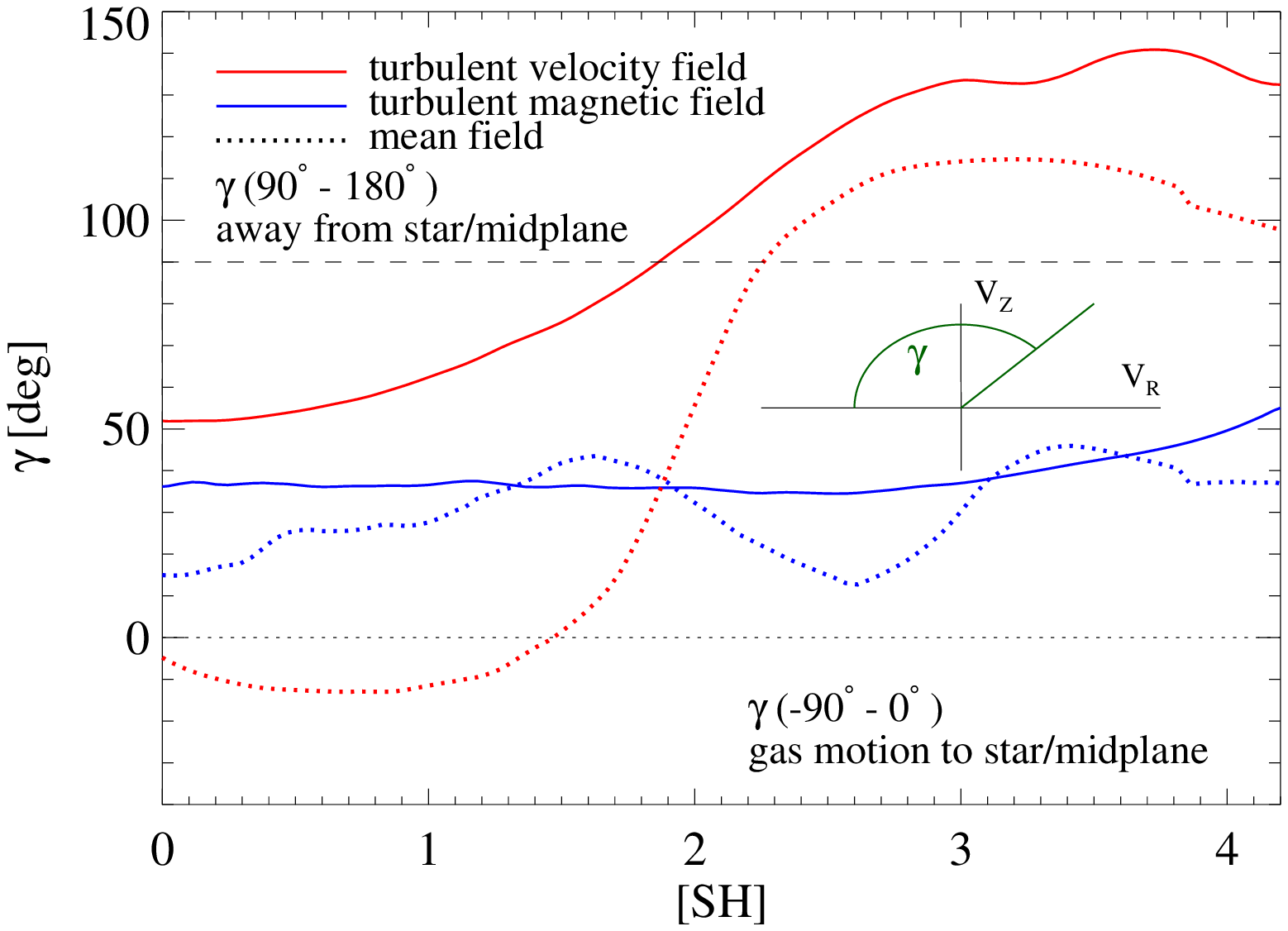,scale=0.56}
\end{minipage}
\hspace{4.2cm}
\begin{minipage}{5cm}
\psfig{figure=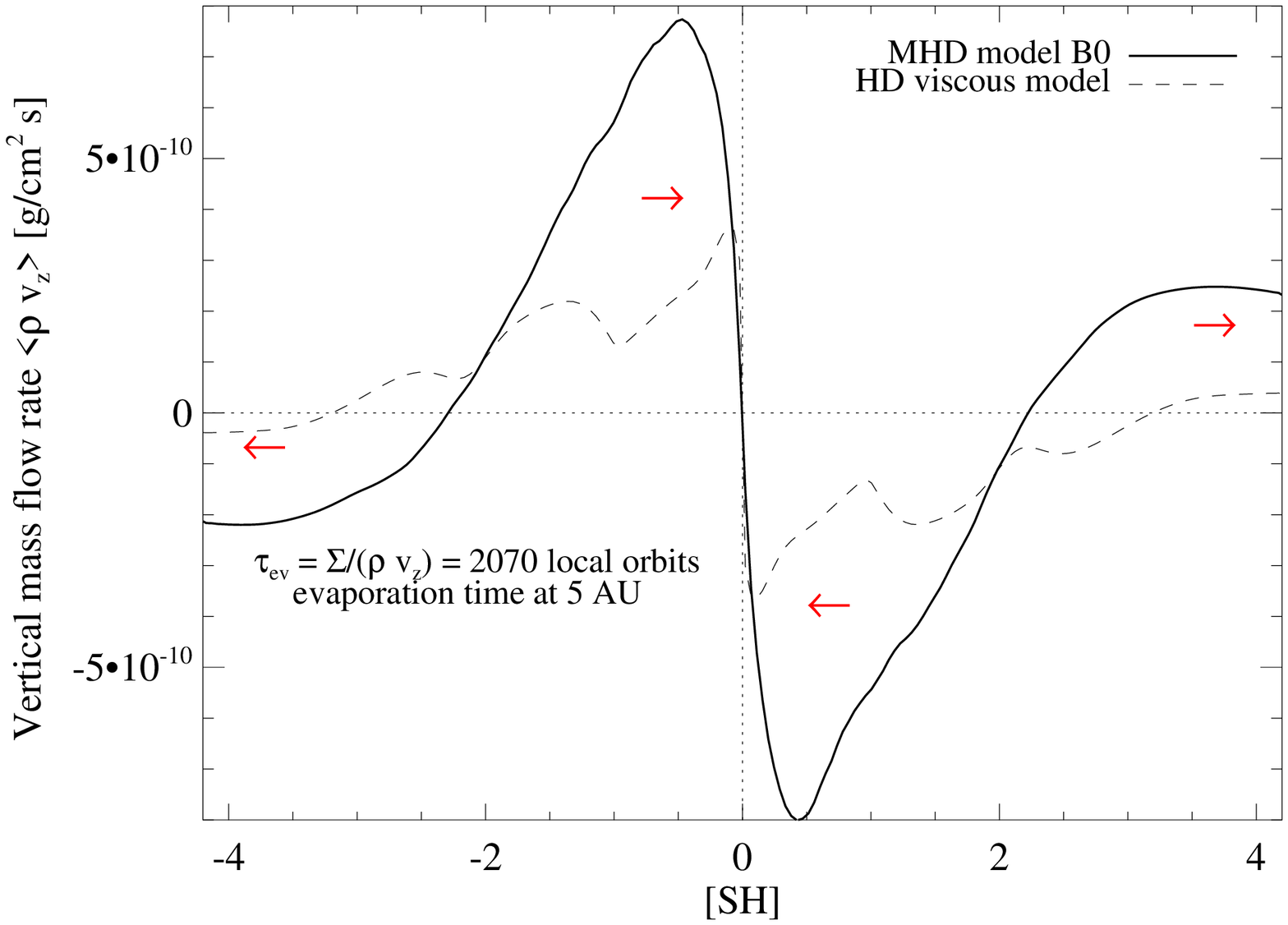,scale=0.46}
\end{minipage}
\label{vel_spec3}
\caption{Left: Angle between the cylindrical radial and vertical velocity 
with respect to the midplane axis ($V_R = -1$ and $V_Z = 0$) for the upper hemisphere.
Right: Vertical mass outflow $\rho v_z dA_z$ in units of $\rm M_\sun/yr$ at 5 AU. 
There is a mass outflow present above 3 scale
heights. The evaporation time, $\rm \tau_{ev} = \Sigma/(\rho v_z)$, was determined to 2070 local orbits.}
\end{figure}

\begin{figure}
\hspace{-1.2cm}
\begin{minipage}{5cm}
\psfig{figure=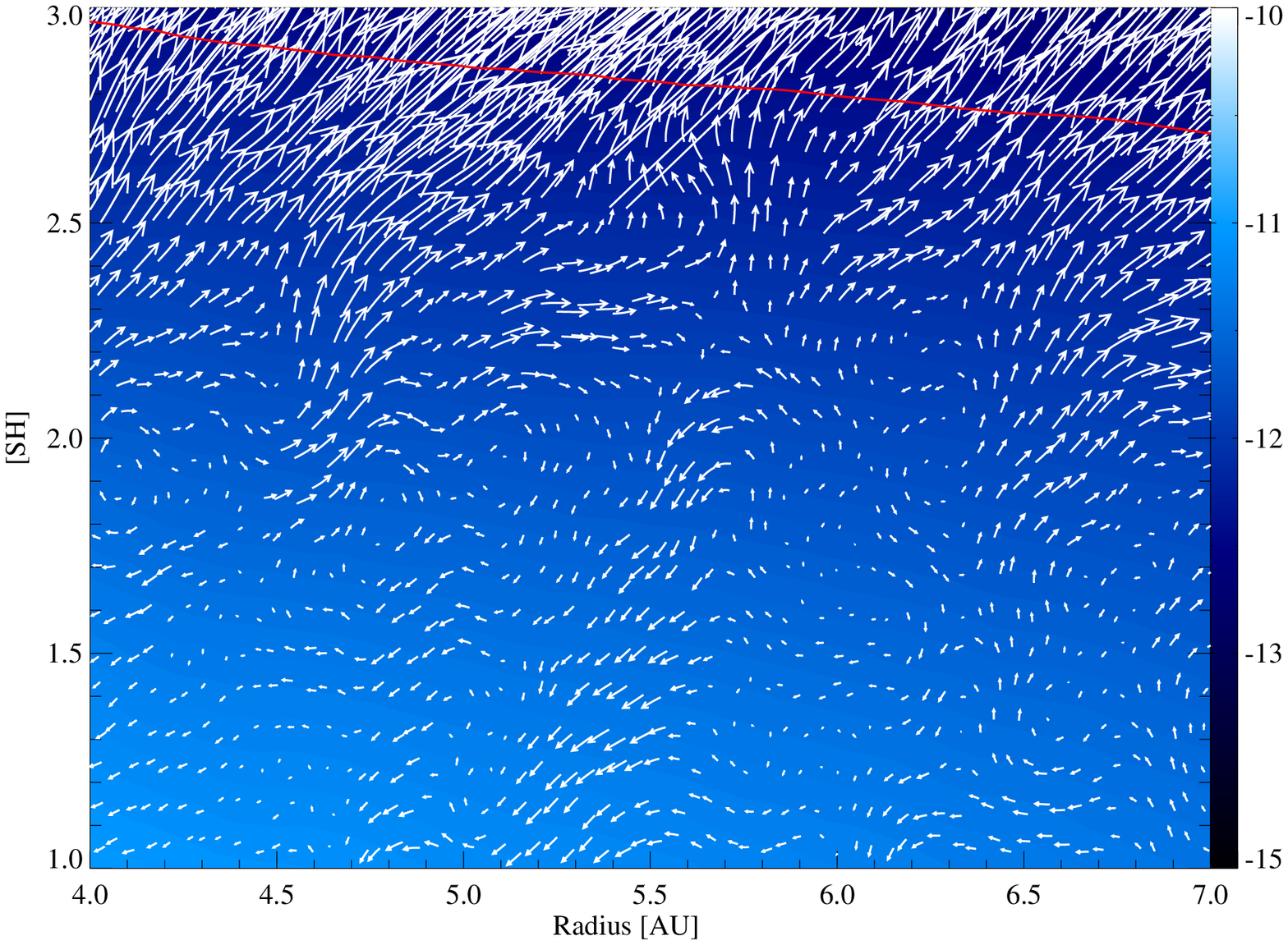,scale=0.49}
\end{minipage}
\hspace{4.2cm}
\begin{minipage}{5cm}
\psfig{figure=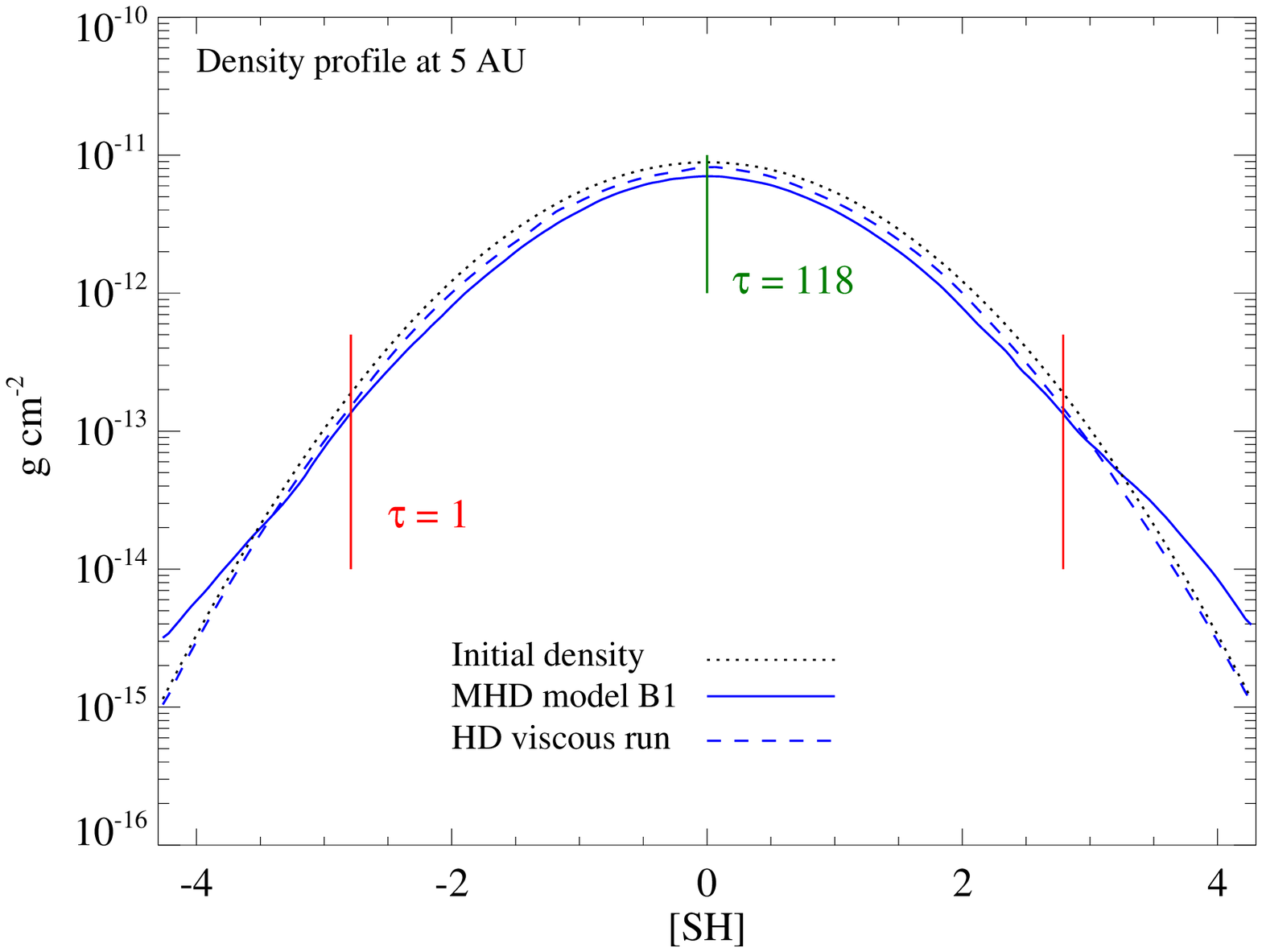,scale=0.46}
\end{minipage}
\label{vel_spec4}
\caption{Left: Logarithmic contour plot of the density, over plotted
with the velocity vector in the $R-\theta$ plane for model BO. Both are averaged over azimuth and time
and plotted for the upper disk hemisphere. The velocity vectors show a
 outflow pattern above two disk scale heights. The red line marks the optical depth of $\tau = 1$ for our setup.
Right: Vertical density profile at 5 AU after 1000 inner orbits for
model BO (solid line), the respective viscous model (dashed line) and the 
initial profile (dotted line). The optical depth of $\tau = 1$ is around 2.8 scale
heights for our model.}
\end{figure}

\begin{figure} 
\hspace{-0.6cm}
\psfig{figure=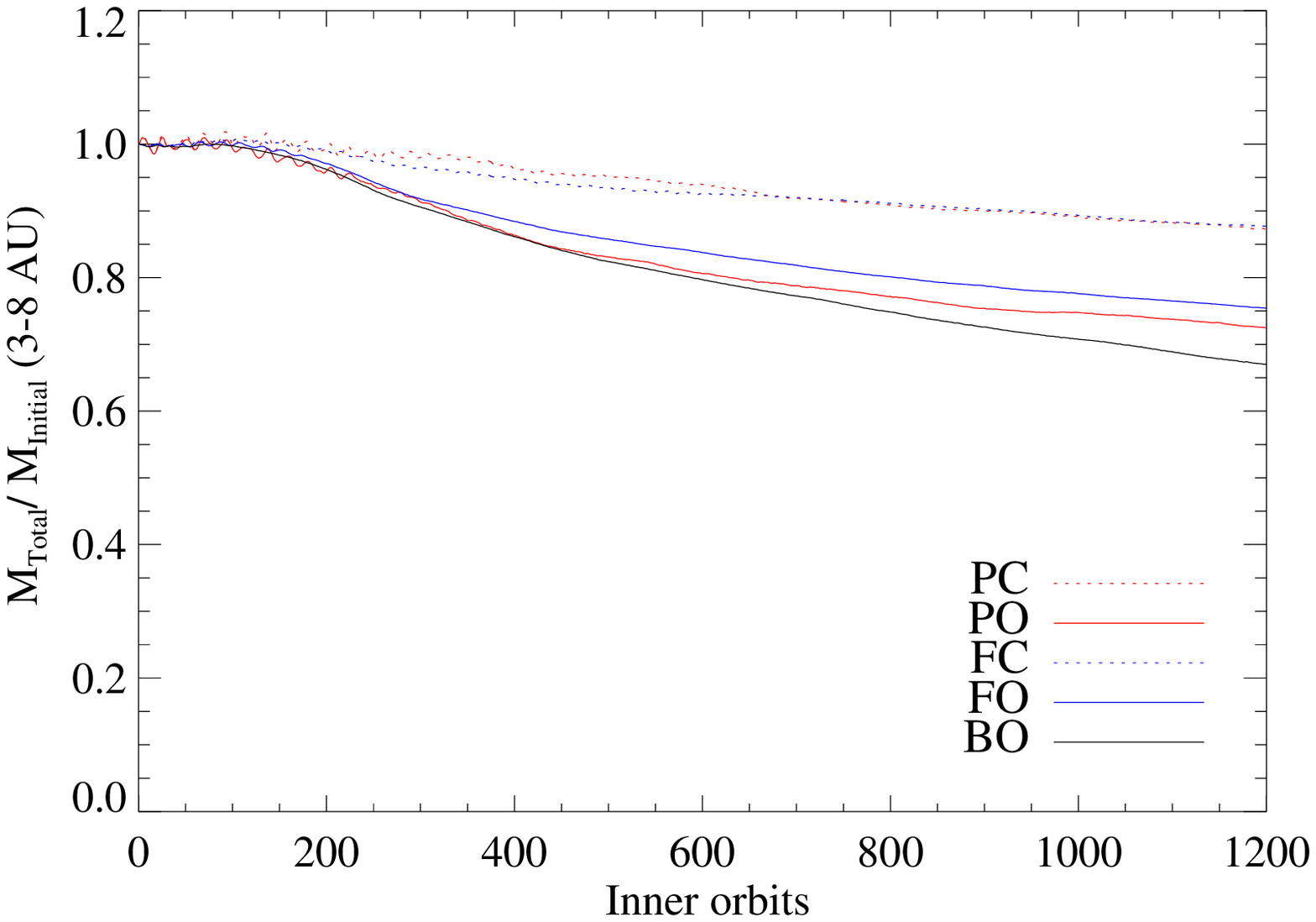,scale=0.46}
\label{totalmass}
\caption{Total mass plotted over time for all models. The mass is integrated in the
space region B (3-8 AU). The change of the mass for the closed models is due
to radial movement. The mass loss for the open models is dominated by the vertical outflow.}
\end{figure}

\subsection{Vertical outflow}
In the previous section, we have seen that the MHD simulations point to the
presence of an additional process besides turbulent "viscous" spreading that removes gas from the disk.
Fig. 4, right, shows the initial vertical density profile at 5 AU (dotted
line), the respective profile for the MHD model BO (solid line) and the viscous HD run (dashed line) after 1000 inner orbits (90 local orbits).
For our model, the vertically integrated optical depth of $\tau = 1$ is around 2.8 scale heights (Fig. 4, red line).
For the calculation of the optical depth, we used Rosseland mean opacities \citep{dra84} with the temperature at 5 AU.
The additional magnetic pressure as well as the vertical mass flow, present in the MHD run, generates higher gas density 
above 3 scale heights compared to the hydrostatic equilibrium.
In Fig. 4, left, we plotted a snapshot of the azimuthally averaged velocity 
field, taken after 1000 inner orbits. The plot indicates a vertical outflow in the disk starting above 
2 scale heights. We measured the angle 
between the cylindrical radial velocity $V_R$ and the vertical velocity
$V_Z$ for the mean and turbulent components (Fig. 3, left). 
The angle is measured with respect to the midplane axis
(pointing to the star, see Fig. 3, left, $V_R = -1$ and $V_Z = 0$).
The upper (red solid line) and the lower hemisphere (blue solid line) 
present similar profiles. 
From the midplane up to $1.8$ scale heights, the turbulent velocity field is directed
upwards but still pointing to the star. 
The low angle of $10^{o}$ for the mean velocities shows the gas motion
pointing to the star and towards the midplane. 
At $1.8$ scale heights the turbulent velocity is pointing away from the
midplane and the star.
Also the mean velocity angle changes quickly in the region between 1.6 and 2 scale heights 
to an outflow angle, e.g., steeper than $90^{o}$. This region coincides with the region where
the vertical outflow is launched \citep{suz10}. 
Above 2 scale heights the angle between the turbulent and the mean velocity
components stays above $90^{o}$, leading a vertical outflow with a small radial outward component.
The so-called dynamical evaporation time is the time to evacuate
the gas completely from the disk assuming no supply of matter. In our
model the value is slightly larger than 2000 local orbits (Fig. 3, right)
which provides a confirmation for the vertical outflow obtained by local box
simulations from \citet{suz10} with a vertical net flux field.

In Fig. 3, right, we plotted the vertical mass flow over height at 5 AU.
The outflow starts at 2 scale heights and reaches mass fluxes of $\rm 10^{-10} M_\sun/yr$ at 5 AU (model BO, solid line). The influence of the
pure outflow boundary is observed to cause the small outflow in the HD viscous run (dashed line).
In the midplane region, the disk reestablishes the hydrostatic 
equilibrium due to the radial mass loss at the midplane (Fig. 6, bottom
right, red solid line). This drives to a small mean vertical motions visible
in the vertical velocity (Fig. 6, bottom right, green dotted line).

The gas does leave the grid with Mach numbers of only
0.5, which is significantly lower that the local escape velocity, which would be about Mach 20.
Even the results indicate a stable vertical outflow, 
without including the sonic point and the Alfv\'enic point in the simulation
it is not possible to make prediction about the flow, leaving or returning to
the disk at larger radii.  
Thus the fate of the vertical outflow to be a disk wind or not will have to be determined in more detail 
in future simulations with a much broader vertically extent. For this study one would
most probably need an additional vertical field in the corona, which could support 
additional propulsion effects like magneto-centrifugal acceleration.
\subsection{Velocity analysis}
Planet  formation processes in circumstellar accretion disk are strongly dependent on the
strength of the turbulence. Turbulence mixes gas and particles, diffuses or concentrates them 
and makes them collide \citep{ilg04,joh05,joh07,bra08,cuz08,car10,bir10}.
The property of MHD turbulence that is important for planet formation are the turbulent velocity
and density fluctuations of the gas.
The density fluctuations are around $10\%$ and follow the results by
\citet{fro06}.
The spatial distribution of the fluctuating and mean part of the velocities 
is presented in Fig. 6.
All results are obtained for time averages from 800 to 1200 inner orbits
and are given in units of the sound speed.
Spatial averaging is performed in azimuth and between 3 and 7 AU for the vertical
profiles. The radial profiles are mass
weighted. 
Fig. 6, top left, shows the turbulent RMS velocity over radius. 
The profile is roughly constant with a total RMS velocity
of $0.1 c_s$, dominated by the radial turbulent velocity.
The vertical dependence of the turbulent velocity (Fig. 6 - right - top) shows a flat profile 
around $\pm 1$ scale height above and below the
midplane for the radial and azimuthal velocity.
Both components increase above one scale height by an order of
magnitude.
The radial component dominates with $0.07 c_{s} $ around the midplane up to $0.3 c_{s} $ at 4 scale
heights. 
The azimuthal component follows with $0.05 c_{s} $ up to $0.2 c_{s}$ at 4 scale heights.
Only the $\theta$-component does not show a flat profile around the
midplane and increases steadily from $0.02 c_{s} $ to $0.2 c_{s}$ at 4 scale heights,
which is an effect of the density stratification.
The small decrease of the $\theta$ component near the vertical boundary is due to the 
outflow boundary because it does not allow inflow velocities.\\

A global picture of the total rms velocity is presented in Fig. 7.
The 3D picture is taken after 750 inner orbits and shows again the 
different turbulent structures of the midplane and coronal region.  
There are also localized supersonic turbulent motions in the disk
corona (Fig. 7, white color). Compared to the turbulent velocity, the mean
velocities of the gas are two
order of magnitude smaller. They show small but steady gas motions in the
disk. 
The vertical dependence for the mean velocity (Fig. 6 right - bottom) 
shows the small inward motion (red solid line) as well as the change of $r$ and $\theta$-velocity
components to an outflow configuration around 1.6 scale heights.
\begin{figure}
\hspace{-0.6cm}
\begin{minipage}{5cm}
\psfig{figure=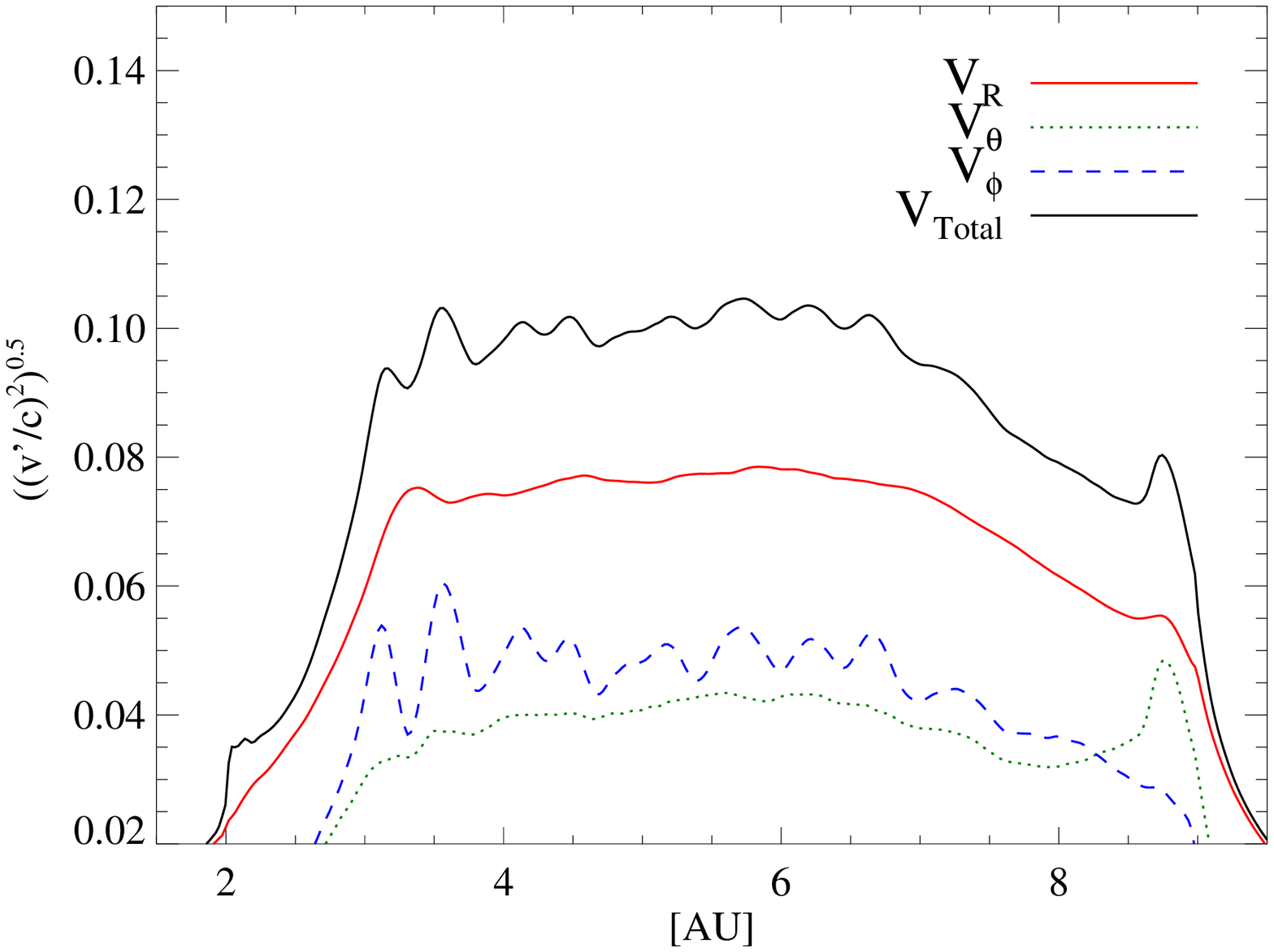,scale=0.46}
\psfig{figure=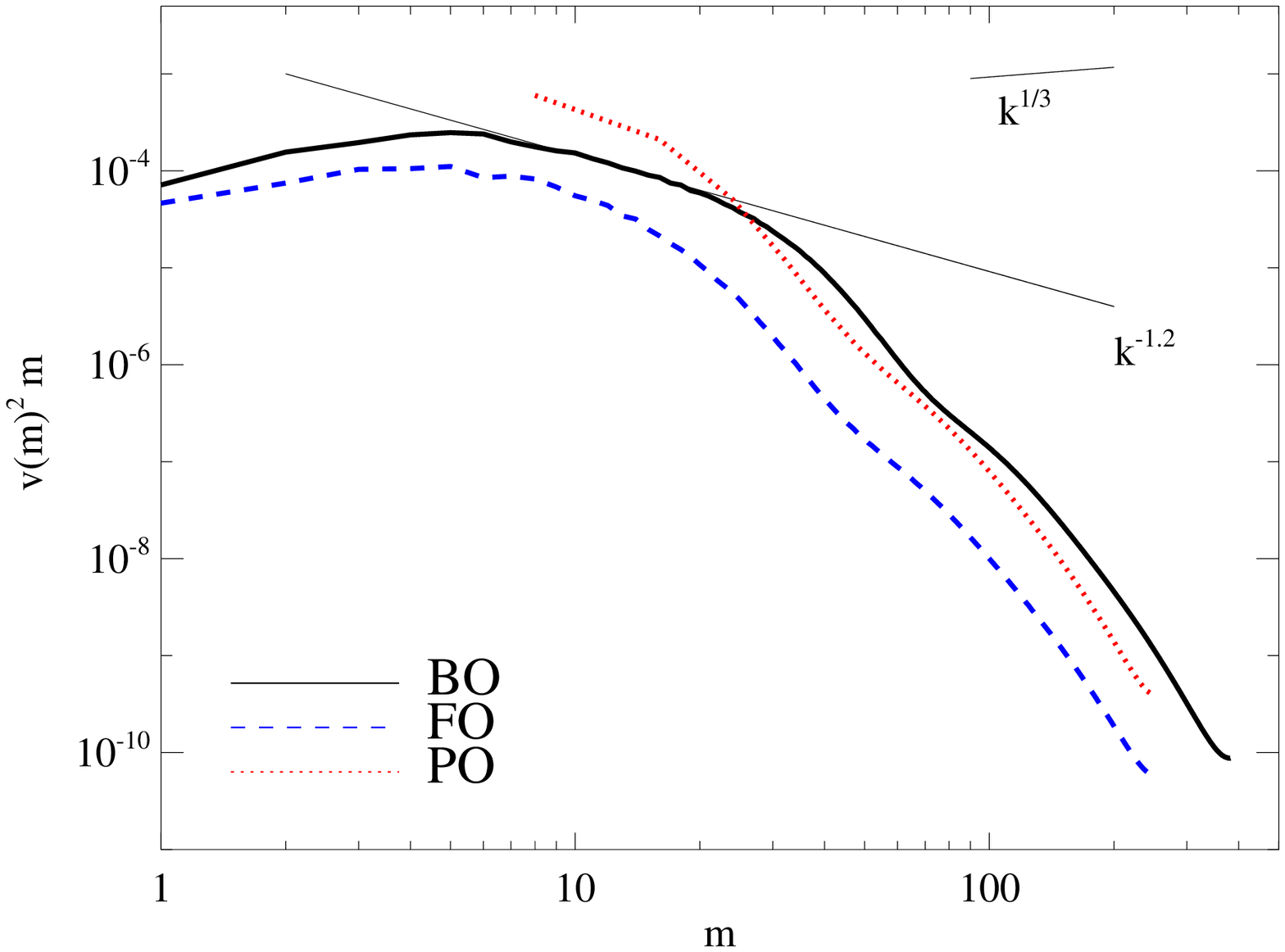,scale=0.46}
\end{minipage}
\hspace{4.0cm}
\begin{minipage}{5cm}
\psfig{figure=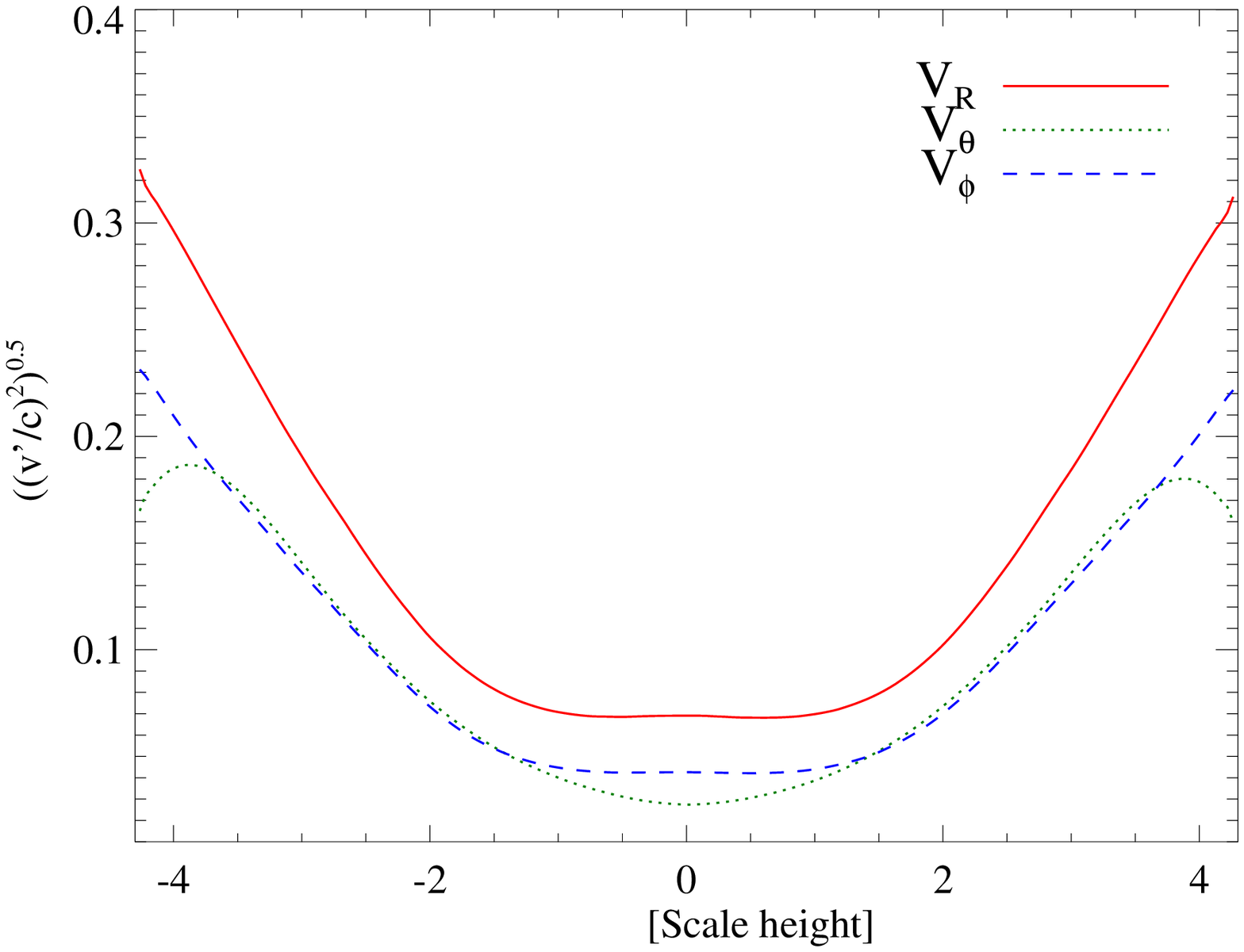,scale=0.46}
\psfig{figure=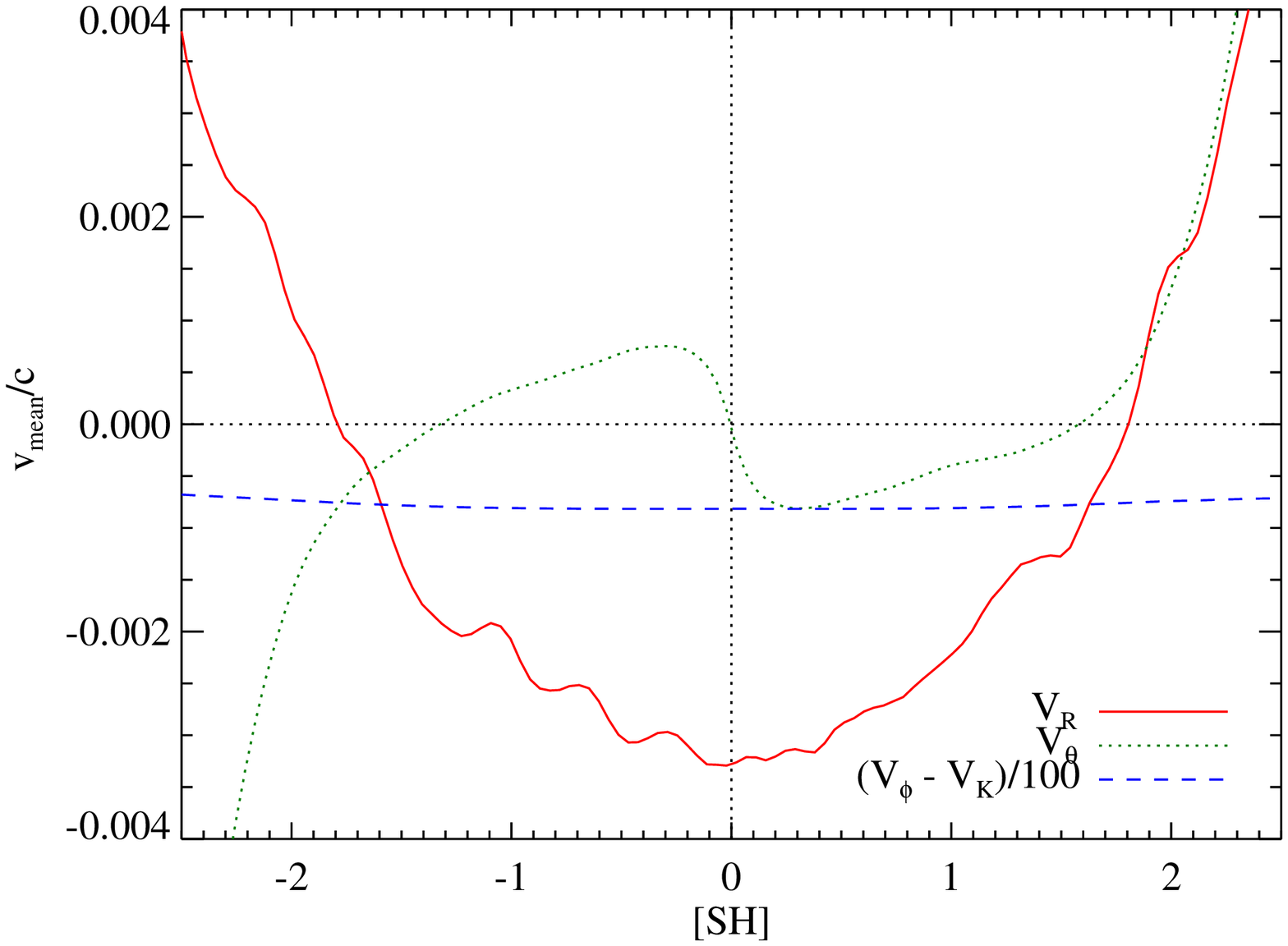,scale=0.46}
\end{minipage}
\label{vel_turb}
\caption{Top left: RMS fluctuations of the velocity versus radius for model BO,
averaged over time and azimuth. 
All components show a roughly flat profile, dominated by the radial turbulent velocity.
The radial profiles are mass weighted. The time average is performed during time period II (Fig.1, top
left, green line).
Top right: Turbulent velocity profile versus scale height for model
BO, averaged over time and azimuth.
There is a flat profile visible in the range $\pm 1.5$
scale heights above and below the midplane.
Starting at $1.5$ scale heights the turbulent velocity increases.
Bottom left: Energy power spectra $E_m\cdot m^2$ for $\pi/4$ model PO (red dotted line),
$2\pi$ model FO (blue dashed line) and the high-resolution model BO (black solid line). 
Bottom right:  Time average of the mean velocity over scale height.
}
\end{figure}

\subsubsection{Kinetic spectra}
Not all particles do couple alike to the turbulent gas flow. 
In fact particles have a size-dependent friction or stopping time \citep{wei77}.
This stopping time is also the time a particle needs to couple to the
turbulent gas flow. 
Best coupled to turbulence are those particles, which have a coupling
time shorter then turbulent correlation time. 
Particle collision velocities are maximized for particles whose stopping time
coincides with the turbulent correlation time, e.g., the eddy turn over time.
This means that particles of different sizes couple to different length scales of the
turbulent spectrum. Therefore, a study of planet formation processes needs 
not only the mean turbulent velocity but also its spectral distribution.\\
In the global domain, only the $k_{\phi}$ space of the spectra is
without modifications accessible as only the $\phi$ direction is periodic in space.
The investigation of the complete $k$-space for this model goes beyond the
scope of this work.
The classical Kolmogorov theory predicts the scaling of the energy
spectra per wavenumber: $E(k) \propto v_k^2 k^{-1} \propto \epsilon^{2/3} k^{-5/3}$. 
We calculate along azimuth $|v(k_\phi)|^2 = |v_r(k_\phi)|^2 +
|v_\theta(k_\phi)|^2 + |v_\phi(k_\phi)|^2 $ with $v_r(k_\phi) = \left
\langle \int_\phi
v_r(r,\theta,\phi)e^{-ik_{\phi}\phi} d\phi \right \rangle$.
The average is done in radius (region B, Fig. 1) and height ($\pm 0.5$ disk scale heights).
For our spectra we use the azimuthal wavenumber $\rm m$ instead of $k$
to be independent from radius: $k = 2\pi/\lambda = m/R$.
In Fig. 6, bottom left, we plotted the energy power spectra $E_m =
v(m)^{2}\cdot m$ with time and
space averaged over $\pm 0.5$ scale heights around the midplane.
In our models we do not observe Kolmogorov inertial like range, $E_m
\cdot m^2 \sim m^{1/3}$.
The $2\pi$ runs F0 and BO have most of the energy placed at $m=5$.
The high resolution model BO present a $k^{-1.2}$ dependence, starting from $m=5$ until
$m=30$.
The $\pi/4$ run PO piles up the energy at its domain size ($m=8$), reaching higher
energy levels compared to the $2\pi$ models FO and BO.
The velocity spectra for each component along the azimuth, 
plotted in Fig. 8, left, indicate that all velocity components have 
similar amplitude for the small scales and do not deviate 
by more than a factor of two at the largest scales.
The radial velocities peak at m equals 5, but 
overall the entire spectrum above m=20 is essentially flat.
The peak at m=5 could be connected to the production of shear waves in the simulations.
These shear or density waves are described in \citet{hei09}.
On top of the shorter time scale of MRI turbulence, these long "time scale" shear waves
are visible in the contour plot of the radial velocity in the $r-\phi$ midplane (Fig. 8 right).
The shear wave structures drive the radial velocity up to $0.3 c_s$.
In the velocity spectra we see the start of the dissipation regime
at $m=30-40$ for the high-resolution run BO. For the model BO, 
this corresponds to 26 or rather 19 grid cells per wavelength, 
which is still well resolved by the code \citep{flo10}.
Shearing waves are also visible in a $r-\theta$ snapshot 
of the velocity (Fig. 9, left). Here we plot the azimuthal 
velocity $V_\phi - V_K$ as contour color, over-plotted with the velocity vectors.
Red contour lines show Keplerian azimuthal velocities.
Super-Keplerian regions are important for dust particle migration.
They reverse the radial migration of particles, leading to their efficient
concentration and triggering parasitic instabilities in the dust layer, like
the streaming instability leading potentially to gravoturbulent planetesimal formation
\citep{kla08,joh07}.
In our simulation these super-Keplerian regions are not completely
axisymmetric, but have
a large extension in the azimuthal direction of several scale heights.
The variation of the orbital frequency over time and space, presented in
Fig. 9, left, and Fig. 2, bottom right, are connected to zonal flows. They
are observed and discussed in several local and global studies
\citep{joh09,dzy10}.
%
\subsection{Magnetic field analysis}
The azimuthal MRI generates a turbulent zero-net field configuration in the
disk.
Despite the loss of mass and magnetic flux through the vertical
boundary there is no sign of decay for the highest resolution case BO 
(Fig. 1, top left, bottom right).
We find well established turbulence. Fig. 9, right, presents a snapshot of the magnetic fields 
after 750 inner orbits. 
The $r-\theta$ components are shown as vectors   
with the azimuthal magnetic field as background color. 
\subsubsection{Magnetic energy spectrum}
To understand the magnetic turbulence at the midplane, we
investigated the spectral distribution of the magnetic energy.
The magnetic energy power spectrum (Fig. 12 - bottom left) is plotted along
the azimuthal direction with the same time and space average as for the 
kinetic energy power spectra.
We plot the magnetic energy power spectra times the wave-number 
$m\cdot B_m^2/2P_{Init-5AU}$ to show where most of the 
magnetic energy is located.
For all runs, most of the magnetic energy is
deposited in small scale magnetic turbulence. This was found in several
recent MRI simulations, latest in local box simulations by \citet{dav10} and \citep{fro10}.
The peak of the magnetic energy lies just above the dissipation regime.
For the $2\pi$ model FO, the peak is located between 
$m=10$ and $m=20$, whereas for the high-resolution run BO this regime
is shifted to $m=20$ and $m=30$. 
The spectra follows closely the $m^{1.0}$ slope until the dissipation regime is reached.
The $\pi/4$ run does not resolve the scales where we observe this $m$ dependence. 
In the restricted model PO most of the magnetic energy is again located 
at the scale of the domain size.\\
\subsubsection{Convergence}
The convergence of MRI is an important aspect in ongoing MRI research in
local and global simulations. In local boxes, there was found convergence for
the large scale turbulence between 32 and 64 grid cells per scale height
\citep{dav10}. Due to the large domain in global simulations, it was up to now not feasible to reach
such resolutions per scale height. Here the first resolution level is needed 
to reach a self sustaining turbulence, at least for
simulations with a zero-net flux toroidal field \citep{fro06}. Comparing the
results from stratified local box simulations we can already give
predictions for global simulations with such high resolutions per scale
height. 
In comparison with the local box simulations by \citet{dav10} we get a very
similar profile of the magnetic energy with increasing resolution.
With higher resolution (FO to BO, Fig. 12, bottom left) the large scale magnetic energy decreases 
while the small scale energy increases. A doubled resolution as model BO should also
show convergence for the large scales. Doing this, we expect
only a weak decrease for the large scale modes, as presented in
\citet{dav10}, Fig. 3.

\subsubsection{Plasma beta}
The overall strength of the magnetic fields is best analyzed by this plasma beta
value $\beta = 2P/B^2$.
Fig. 10 presents a 3D picture of the logarithmic plasma beta for the 
$r-\theta$ components, taken at 750 inner orbits. 
The two-phase structure of the disk is again visible.
The well established turbulence at the midplane has a broad distribution 
of high plasma beta values (Fig. 11 , top left).
In contrast, there are regions in the corona of the disk with plasma beta below unity (Fig. 10, black regions).
The azimuthal and time averaged plasma beta at the midplane lies around 400 (Fig. 11, bottom right).
In Fig. 11, top left, we plot the correlation of plasma beta over height in a scatter plot of all grid cells.
We find the distribution of beta values to be very narrow in the disk corona (1-10) but on the
other hand to be much broader (10 - $10^4$) around the midplane, but strongly peaked around $\beta$ = 500. 
The value of plasma beta in the disk corona depends on several issues. A
zero-net flux MRI turbulence with toroidal field produces lower magnetic
fields in the corona. This was already shown in a very similar simulation by \citet{fro06} (Fig. 8, solid line, model S2). 
In contrast, a vertical initial field produces a stronger turbulence level
with plasma beta values below unity in the corona.
The boundary condition also affects the values in the corona.
A closed boundary condition, e.g. periodic in the vertical direction 
will accumulate large amount of magnetic flux in the corona and drive to a 
plasma beta value smaller then one (observed in model FC and PC). 
The small increase of plasma beta above 3 disk scale
heights is connected to the vertical outflow and the increase of gas
pressure and density in this area (Fig. 4, right). This effect has to be investigated in future work
with a much broader vertical extent.
Very high plasma beta values in the midplane (Fig. 11, top left) indicate reconnection.
Two magnetic fields with different sign and comparable strength coming
too close to each other, e.g., in the same grid cell, do reconnect. 
Such reconnections are visible in single grid cells with nearly no magnetic field.
For our BO model, the reconnection zones reach plasma beta values up to $10^{11}$. 
The heating due to reconnection in those regions is not covered in our
isothermal model, but shall be a subject for future studies.

\begin{figure}
\begin{minipage}{5cm}
\psfig{figure=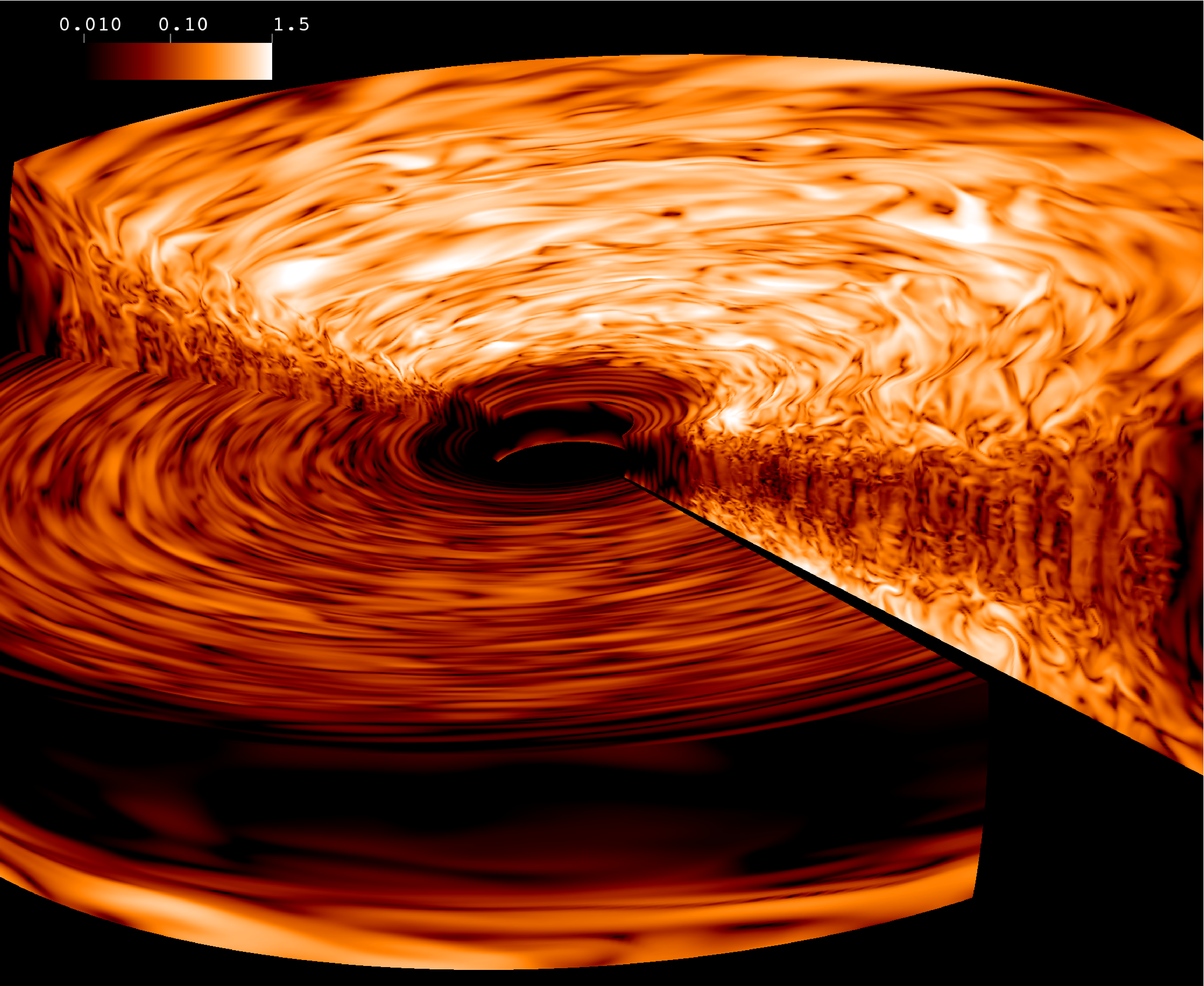,scale=0.22}
\end{minipage}
\label{vrm}
\caption{3D picture of turbulent RMS velocity at 750 inner orbits for model BO.
The white regions in the corona present super sonic turbulence.}
\end{figure}
\begin{figure}
\hspace{-0.6cm}
\begin{minipage}{5cm}
\psfig{figure=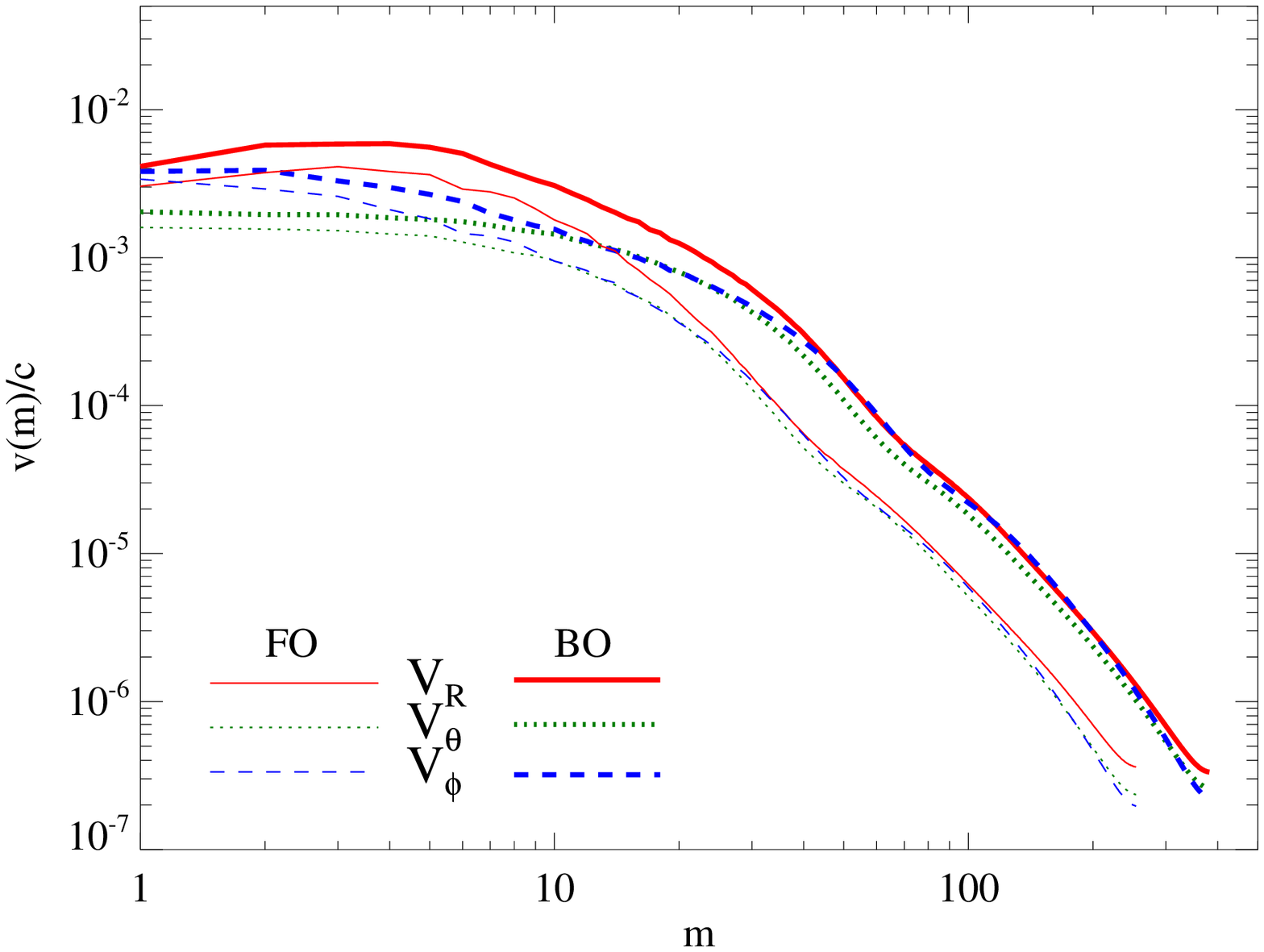,scale=0.46}
\end{minipage}
\hspace{4.0cm}
\begin{minipage}{5cm}
\psfig{figure=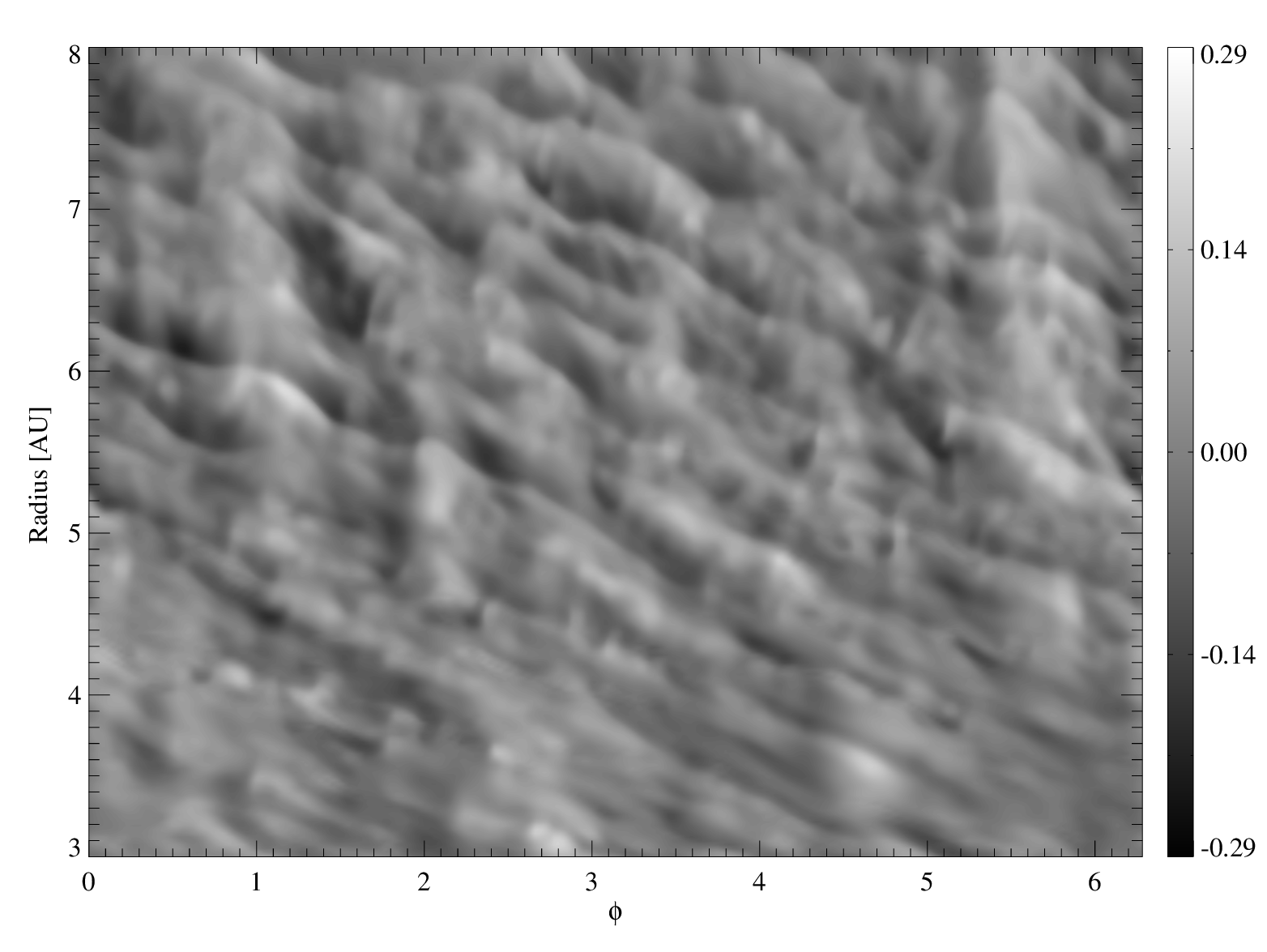,scale=0.46}
\end{minipage}
\label{vel_spec2}
\caption{Left: Velocity spectra in units of the sound speed for all three components
at the midplane. Space and time averaged is again between 3 and 8 AU
and between 800 and 1200 inner orbits. The radial velocity peaks at $m=3-5$
for both $2\pi$ models.
Right: Contour plot of the radial velocity at the midplane ($R-\phi$ plane).
Large shear wave structures become visible. This snapshot is taken after 750
inner orbits.}
\end{figure}
\begin{figure}
\hspace{-0.6cm}
\begin{minipage}{5cm}
\psfig{figure=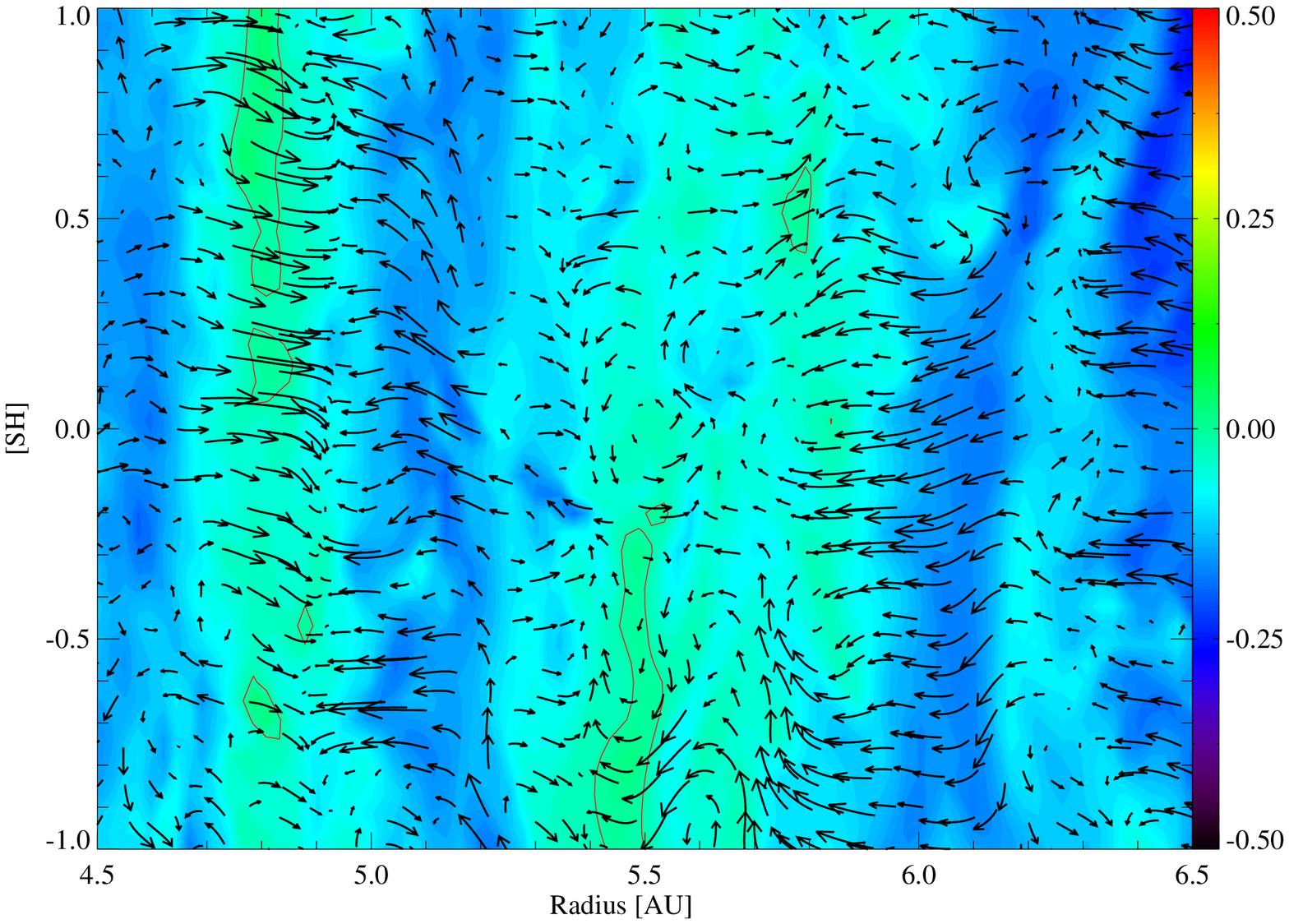,scale=0.46}
\end{minipage}
\hspace{4.0cm}
\begin{minipage}{5cm}
\psfig{figure=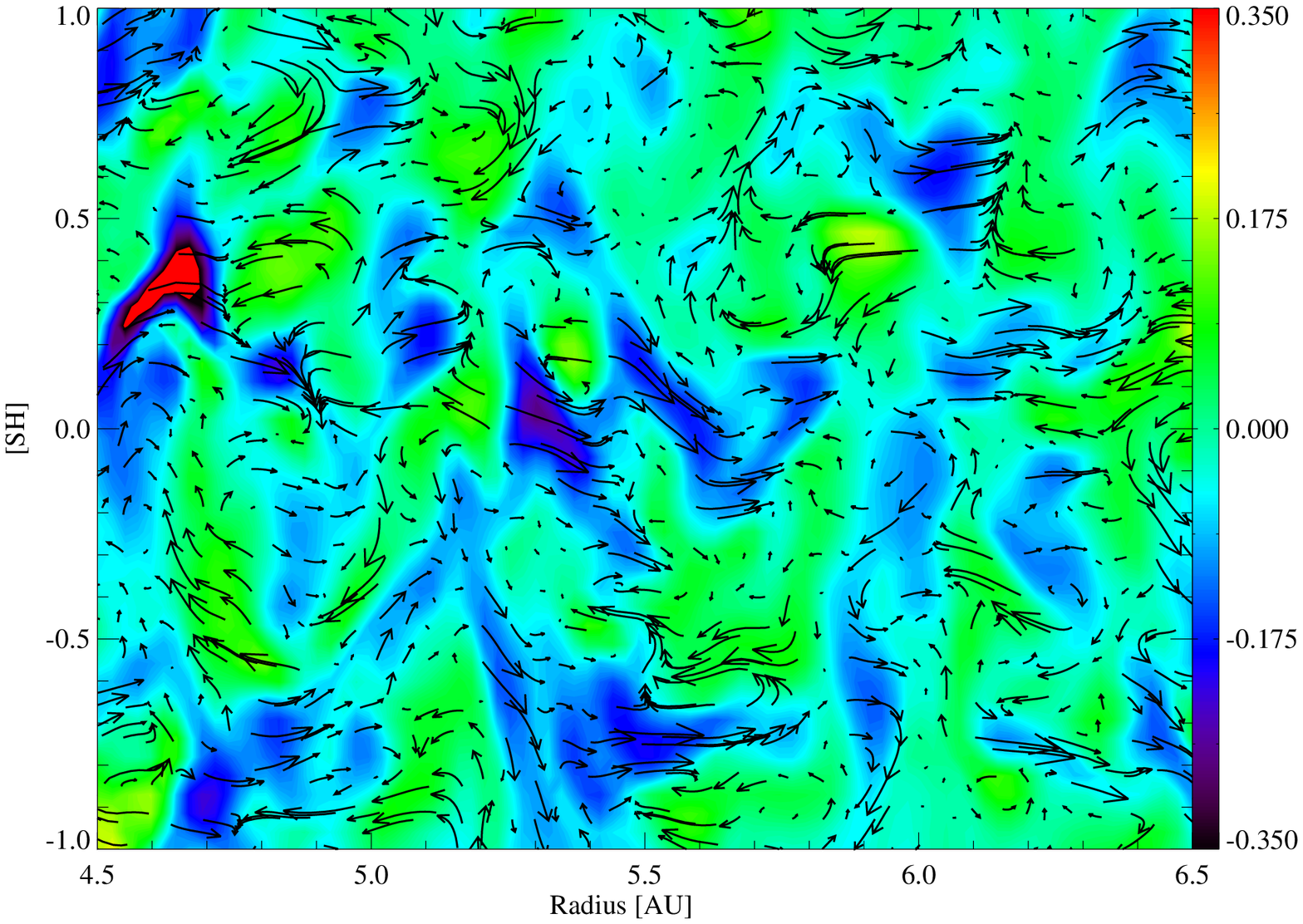,scale=0.46}
\end{minipage}
\label{pow_spec}
\caption{Left: Contour plot of $V_\phi - V_K$ for an azimuthal slice. The red
contour line encloses regions with Super-Keplerian velocity.
Overplotted are the $r-\theta$ velocity fields.
Right:  Contour plot of $B_\phi$ for an azimuthal slice.
Overplotted are the $r-\theta$ magnetic fields fields.
Both snapshots are taken after 750 inner orbits.
}
\end{figure}
\begin{figure}
\begin{minipage}{5cm}
\psfig{figure=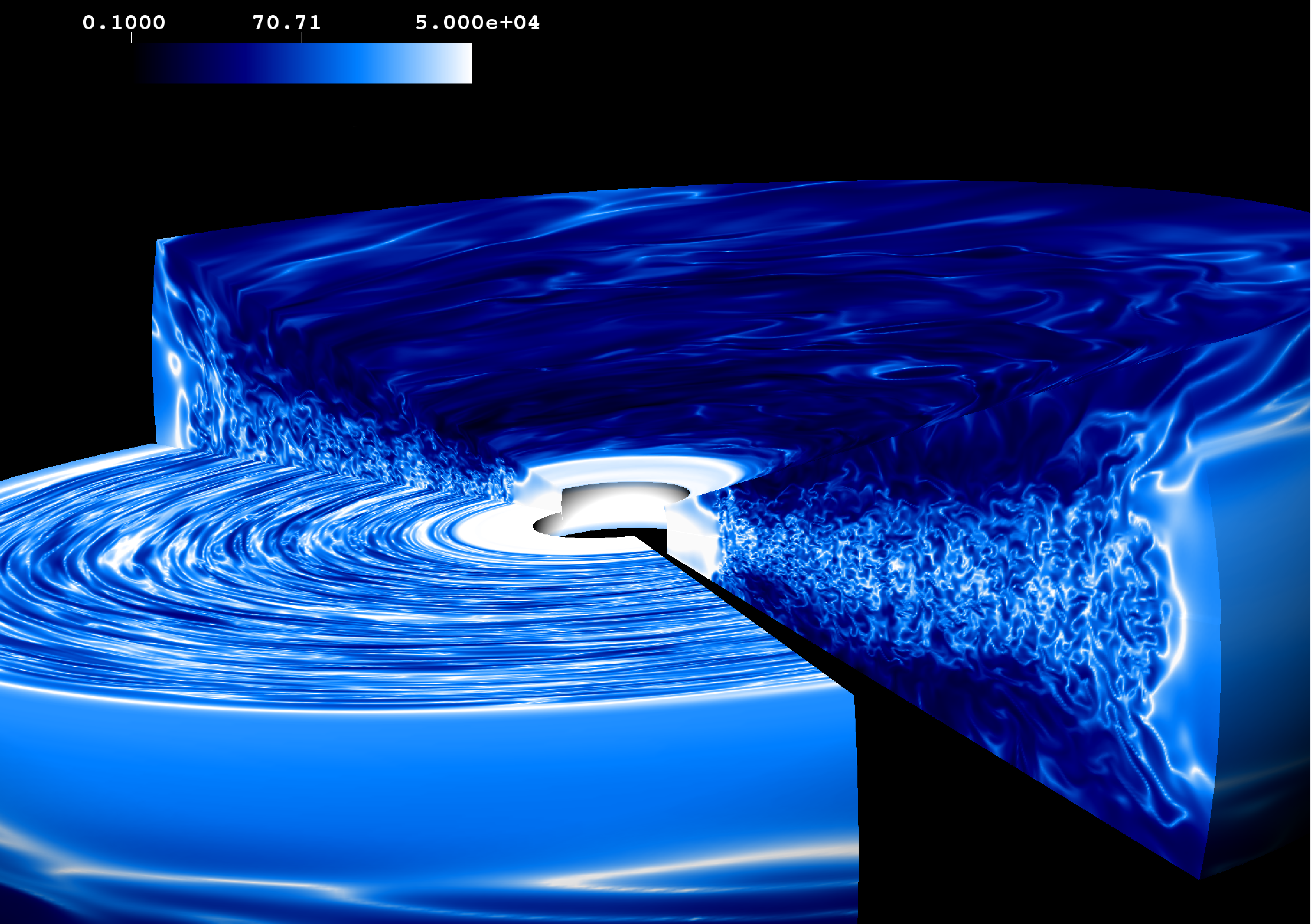,scale=0.22}
\end{minipage}
\label{pbeta}
\caption{3D picture of plasma beta after 750 inner orbits for model BO.
The black regions in the corona present plasma beta values below unity.}
\end{figure}
\begin{figure}
\hspace{-0.6cm}
\begin{minipage}{5cm}
\psfig{figure=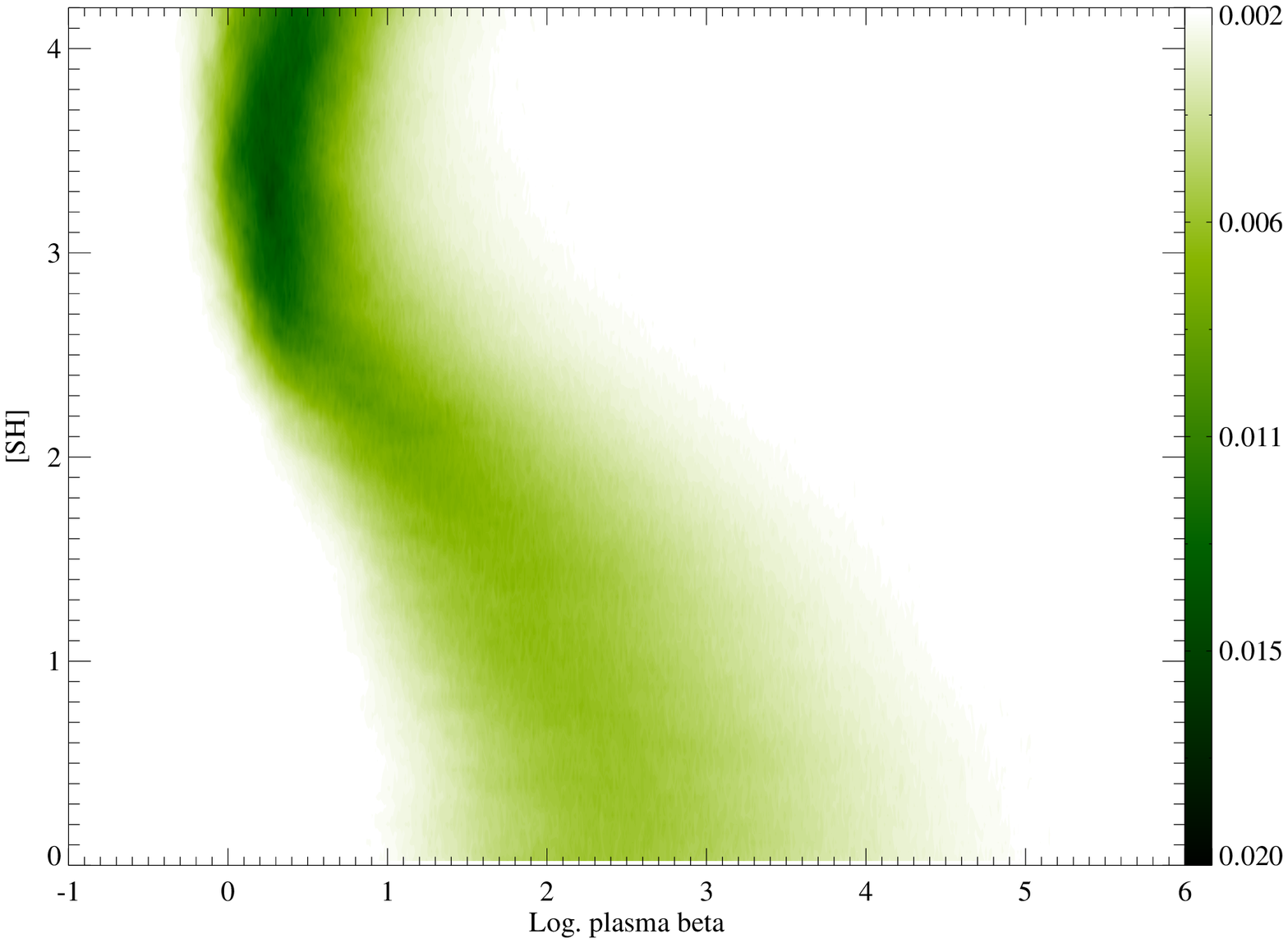,scale=0.46}
\psfig{figure=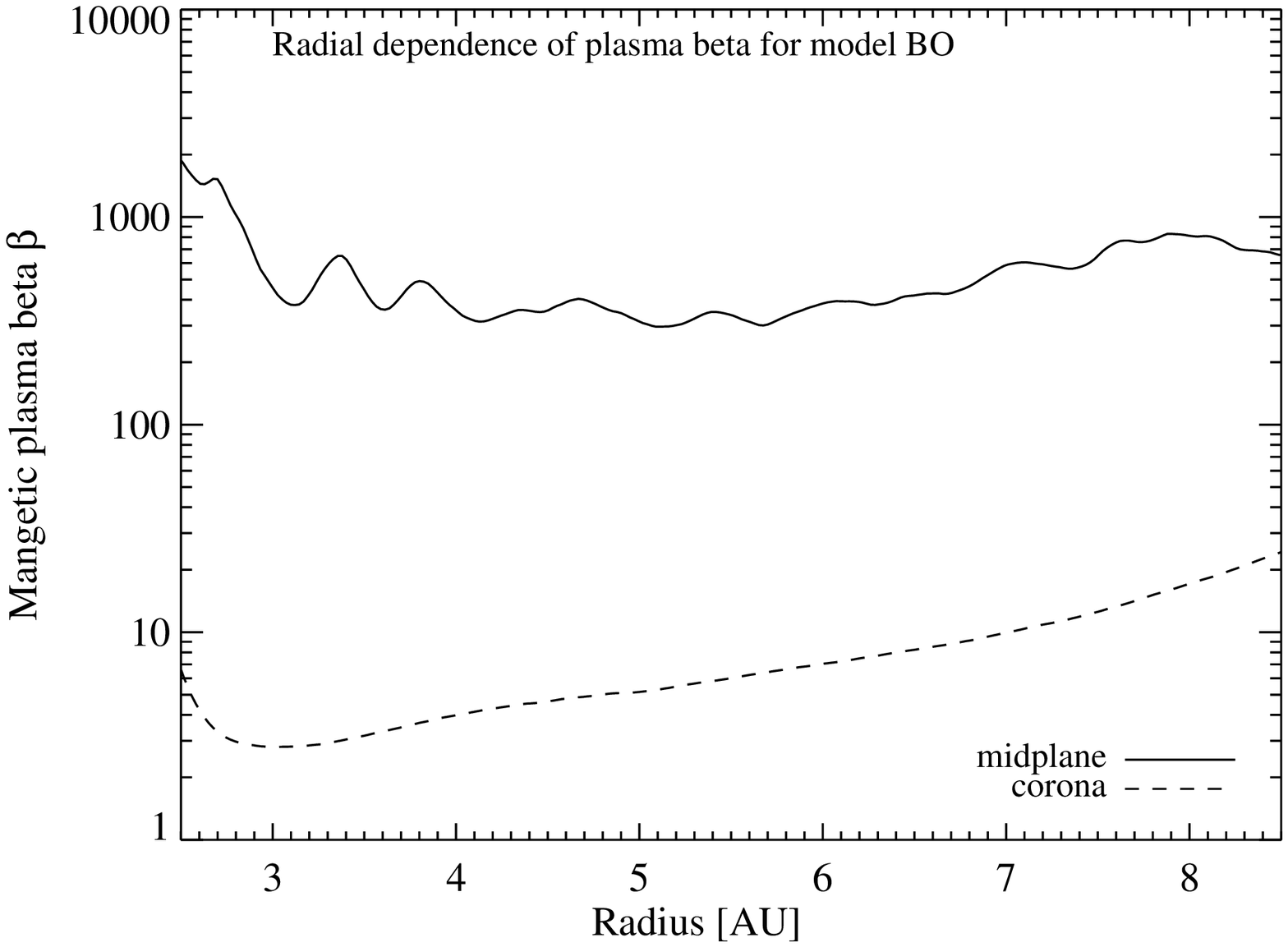,scale=0.46}
\end{minipage}
\hspace{4.0cm}
\begin{minipage}{5cm}
\psfig{figure=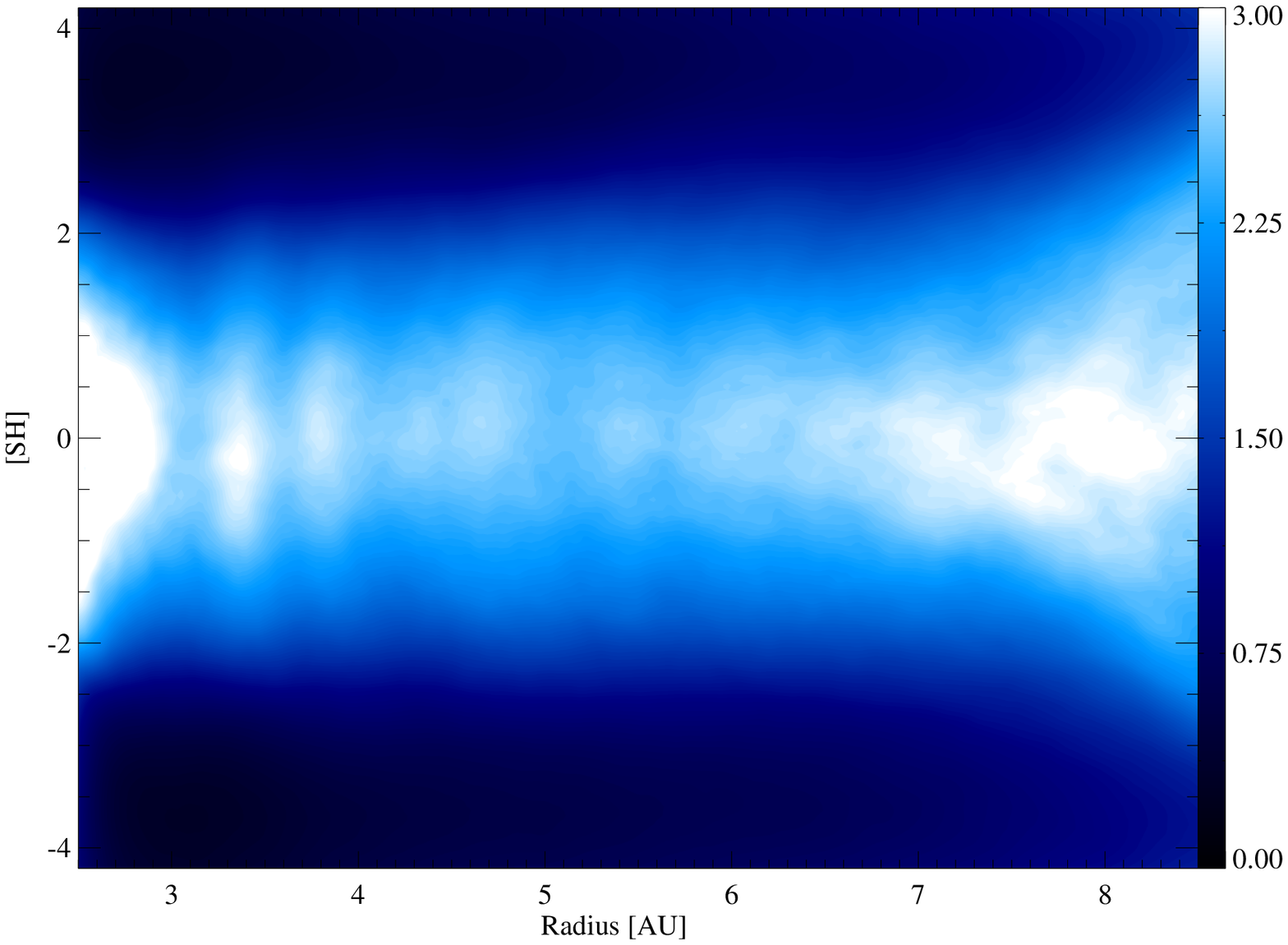,scale=0.46}
\psfig{figure=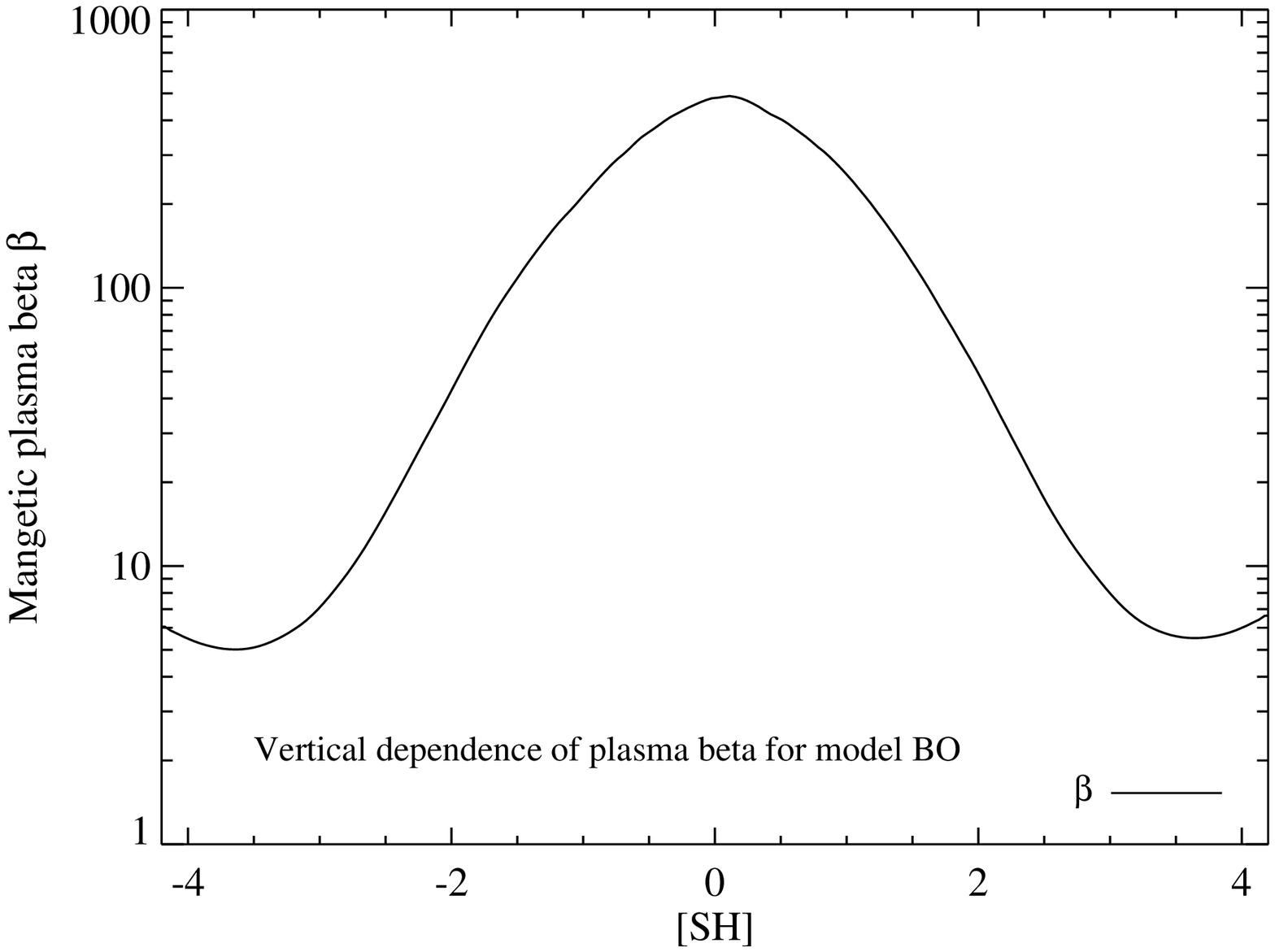,scale=0.46}
\end{minipage}
\label{mag_turb2}
\caption{Top left: Distribution of plasma beta, $N(\beta)/N_{Total}$, over height at 750 inner orbits for model BO.
The color represents the relative number of grid cells, containing specific
plasma beta values. At the midplane,
there is a wide distribution of plasma beta values between 10 and 10000.
In the coronal region the distribution becomes more narrow with values between 1 and 10.
Top right: Contour plot of azimuthal and time averaged plasma beta of BO
with radial (bottom left) and vertical profile (bottom right).}
\end{figure}
\subsubsection{Spatial distribution}
As we already mentioned in section 3.2, the radial profile of the turbulent magnetic field
has a direct effect onto the radial profile of the Maxwell stress in the $\alpha$ parameter.
The dominant turbulent azimuthal magnetic field goes as $1/r$, as shown in a
azimuthal and time average in Fig. 12, top left.
The saturated turbulent field is 4 times lower than the initial azimuthal field.
All values are normalized to the initial gas pressure at 5 AU at the midplane
and the radial profiles are again mass weighted.
The vertical profile shows a constant distribution around $\pm 2$ scale heights 
from the midplane until it decreases with height (Fig. 12, top right).
In contrast, the radial and $\theta$ component show a local
minimum at the midplane with a peak of turbulent magnetic field slightly
above 2 scale heights. \\
The turbulent magnetic fields are around 2 orders of magnitude larger than the
mean fields. 
The vertical profiles of mean magnetic fields over height are presented in Fig. 12, bottom right.
The radial magnetic field is anti-symmetric to the midplane and correlated
with the dominating azimuthal component.
The distribution of mean magnetic fields are connected to the "butterfly"
oscillations.

\subsubsection{Butterfly structure}
The butterfly pattern is a general property of MRI turbulence and  
was found in many local and global simulations, latest by
\citet{gre10}, \citet{fla10}, \citet{sor10} and \citet{dzy10} .
The "butterfly" pattern becomes visible for the mean $B_\phi$ evolution, 
plotted over disk height and time.
In Fig. 13, bottom, we plotted the $B_\phi$ component of the magnetic field
averaged over a small radius (4 - 5 AU) and over azimuth for model FO, left,
and PO, right. 
We see a clear "butterfly" pattern in both models. 
This pattern is also visible in the total accretion stress with
doubled period (Fig. 1, bottom right). In comparison, the $\pi/4$ run shows 
no systematic and more violent picture of the butterfly. The amplitudes are stronger and 
it has mixed symmetry (Fig. 13, bottom right). Also the total magnetic flux
evolution shows these properties for model PO (Fig. 13, top right). 
The FO run presents a similar amplitude and period as the BO run. 
The effect of the narrow azimuthal domain on the mean fields will be investigated in a follow-up work.
The reason of this butterfly structure and its role for the MRI is still under discussion. 
Recent studies show the connection to the MHD dynamo \citep{gre10}
and magnetic buoyancy \citep{shi10}.

\begin{figure}
\hspace{-0.6cm}
\begin{minipage}{5cm}
\psfig{figure=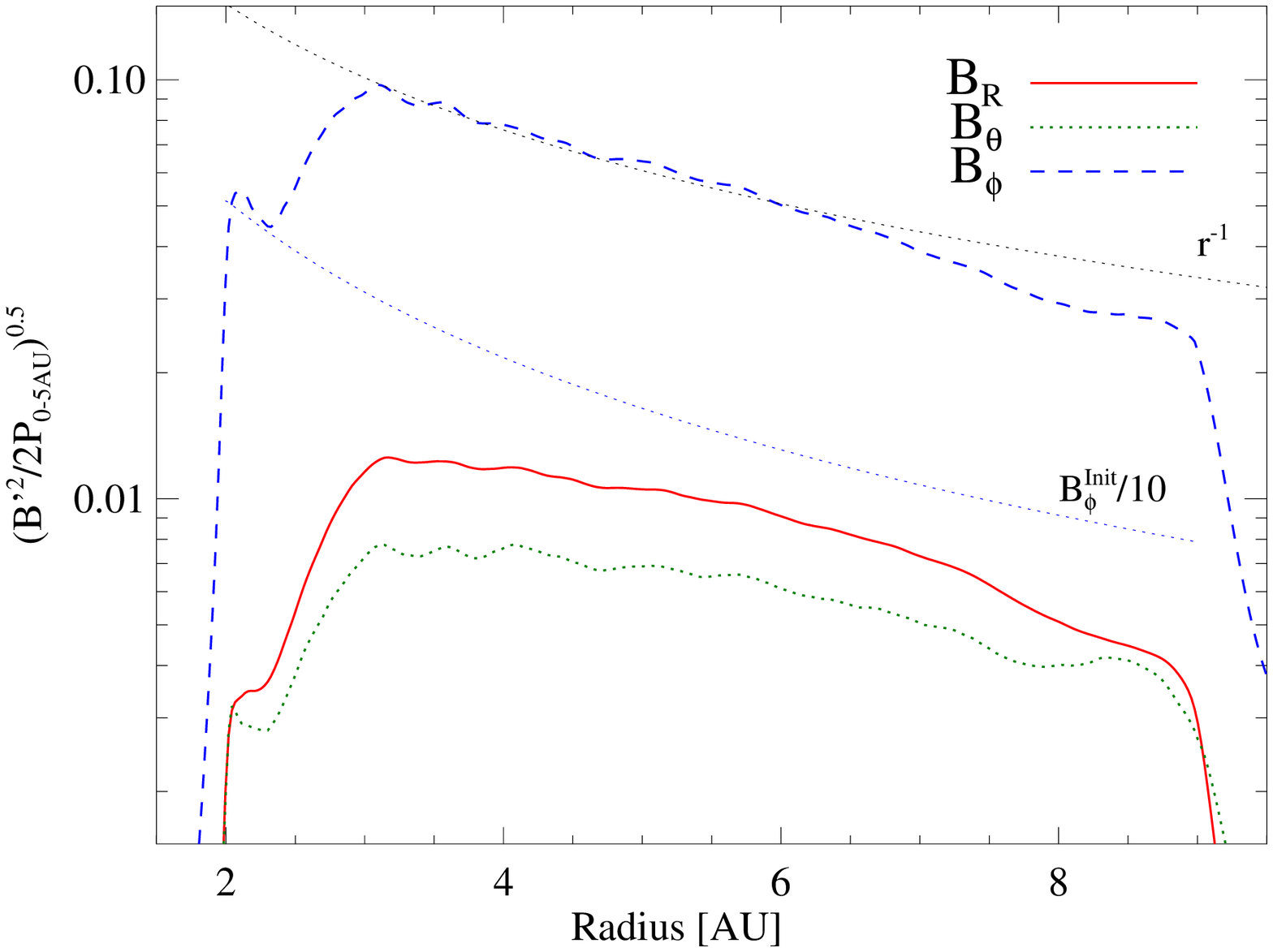,scale=0.46}
\psfig{figure=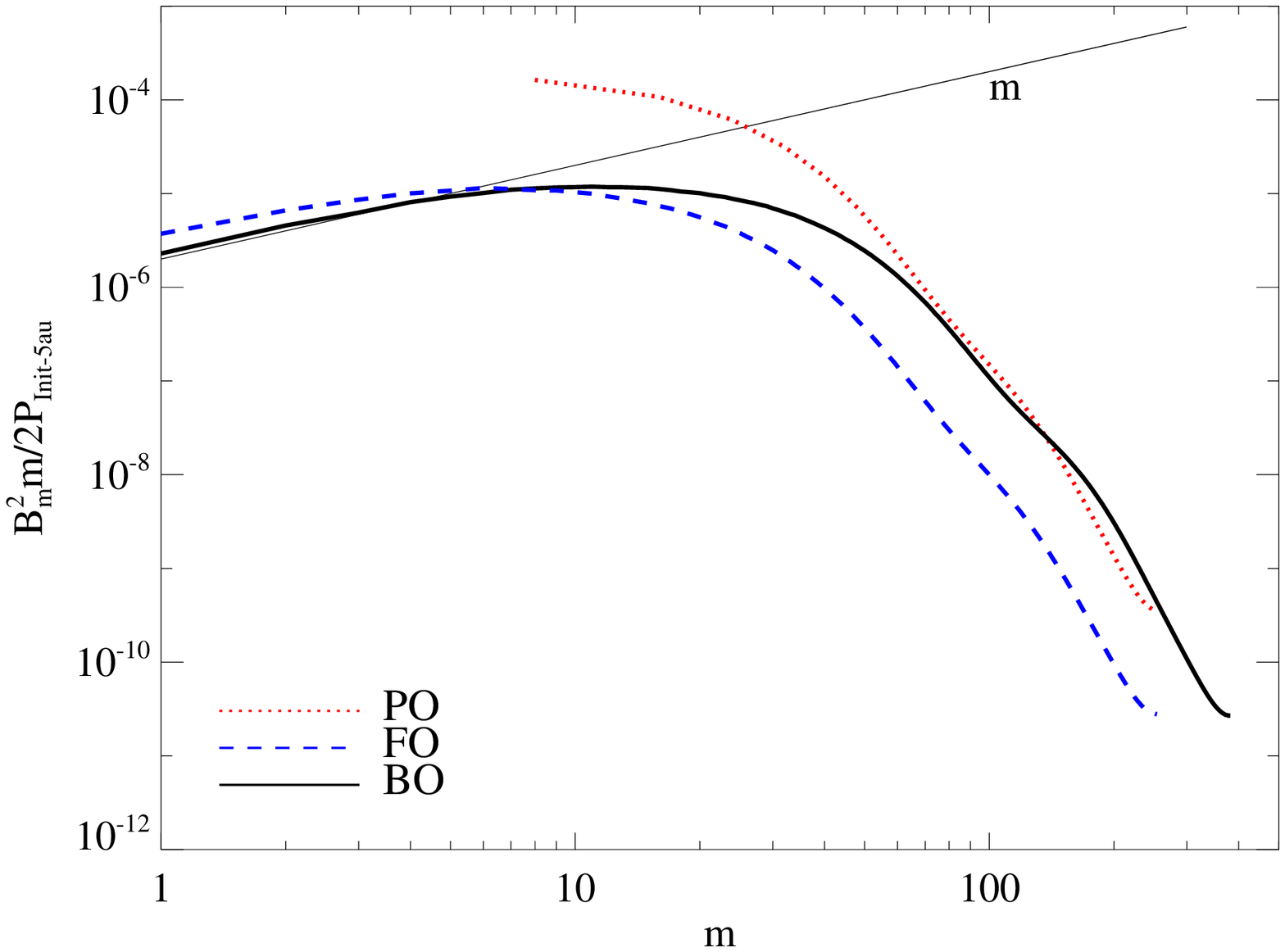,scale=0.46}
\end{minipage}
\hspace{4.0cm}
\begin{minipage}{5cm}
\psfig{figure=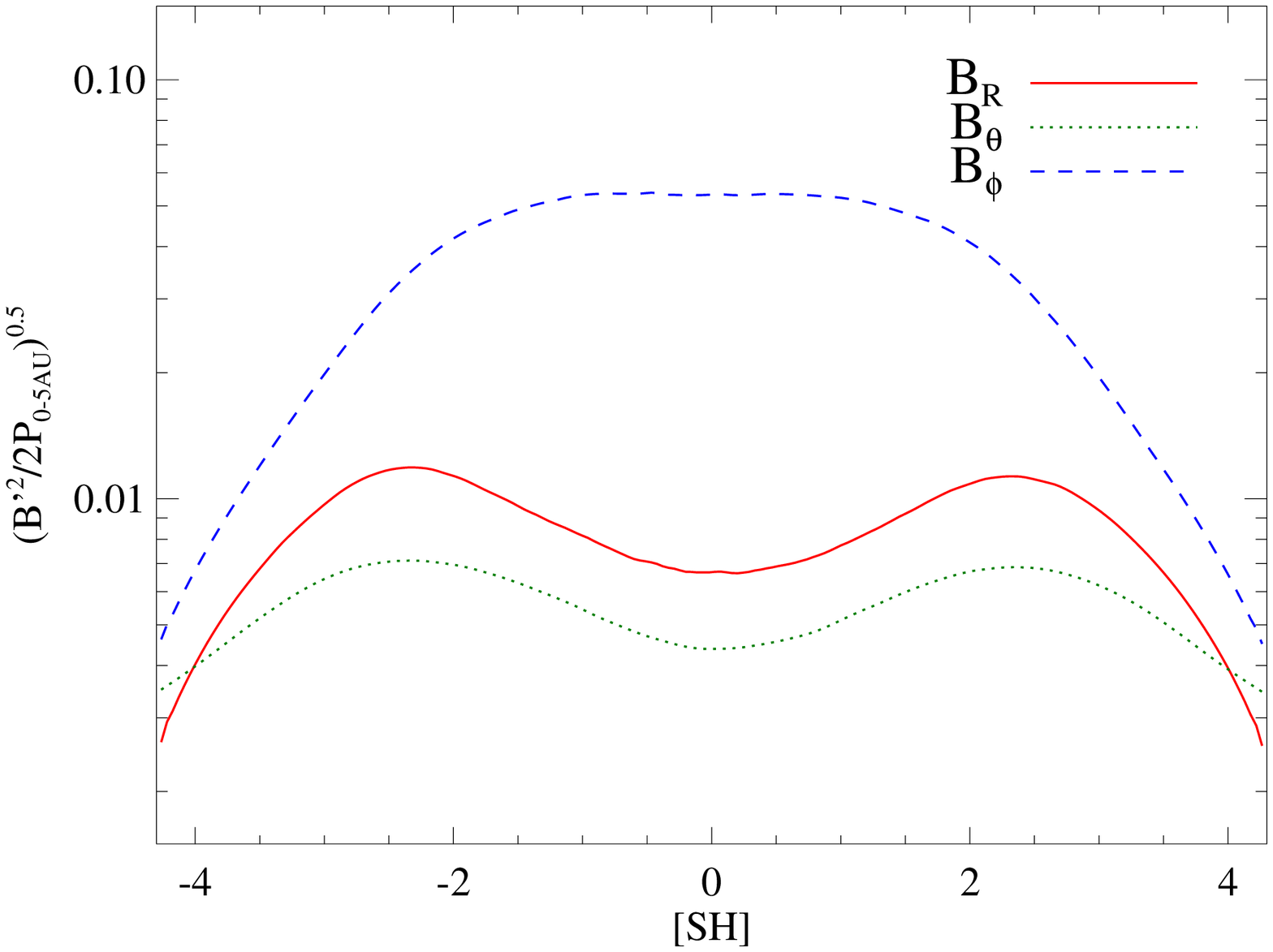,scale=0.46}
\psfig{figure=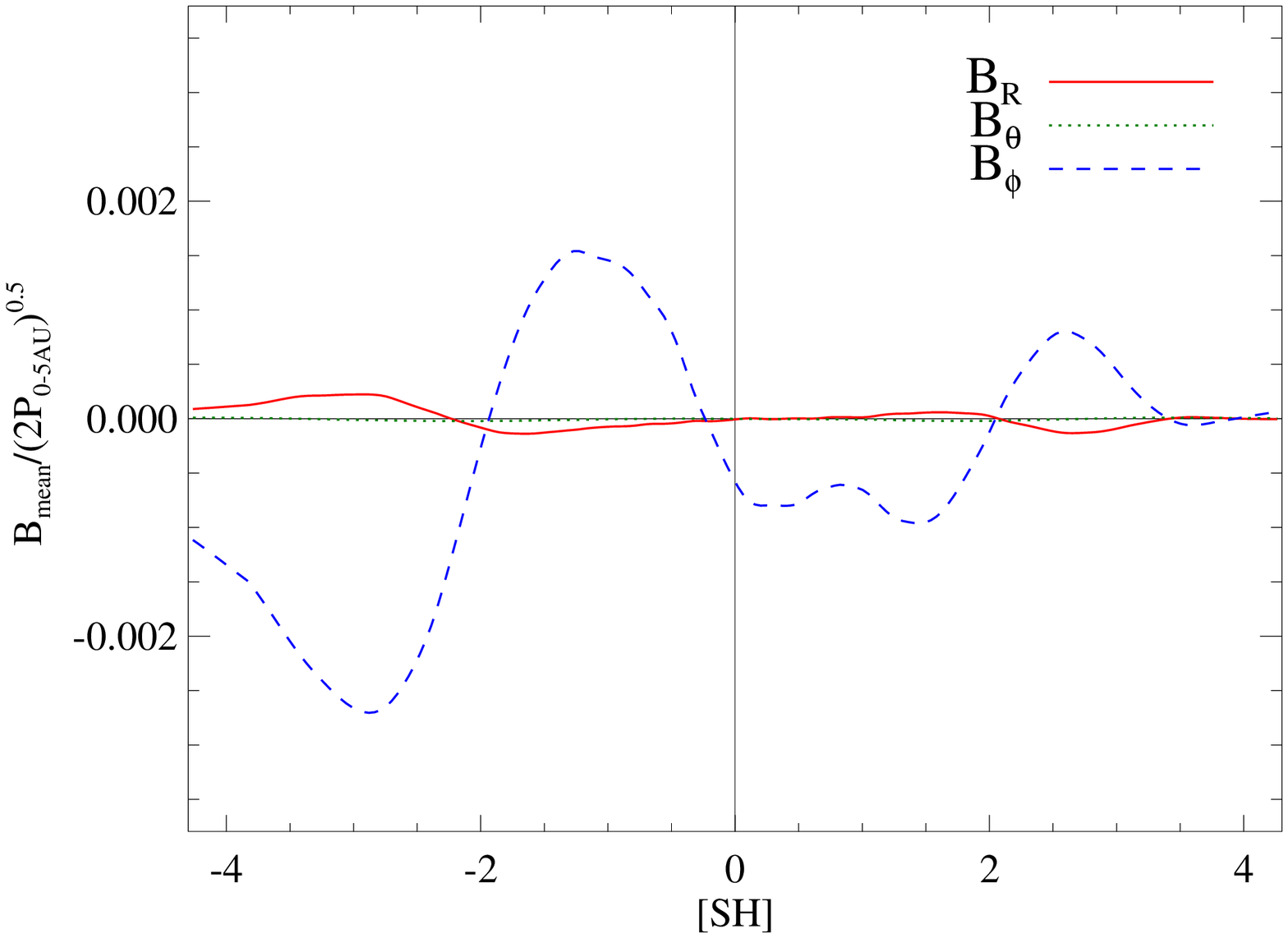,scale=0.46}
\end{minipage}
\label{mag_turb}
\caption{Top left: Time averaged turbulent magnetic field over radius for
model BO. The turbulent field adjusts to the force-free $r^{-1}$ profile.
Top right: Time-averaged turbulent magnetic field over scale height for
model BO. The dominating turbulent azimuthal field represents the same flat profile
$\pm 1.5$ scale heights around the midplane as the velocity (Fig. 6, top
right). The turbulent radial and $\theta$ components represent a different profile with
maximum at 2.3 scale heights.
Bottom left:  Magnetic energy power spectra $B_m^2\cdot m$ for $\pi/4$ model PO (red dotted line), $2\pi$ model FO
(blue dashed line) and the high-resolution model BO (black solid line).
The profile follows the $m^{1.0}$ slope until the dissipation range.
Bottom right:  Time-averaged mean magnetic field over height for     
model BO. The radial and azimuthal field show again anti-correlation.
The anti-symmetry for the upper and lower hemisphere could be correlated
with a $\alpha$-$\Omega$ MHD dynamo.
All radial profiles are mass weighted. The time averaged is performed in time
period II (Fig.1, top left, green line).}
\end{figure}
\begin{figure} 
\hspace{-0.6cm}
\begin{minipage}{5cm}
\psfig{figure=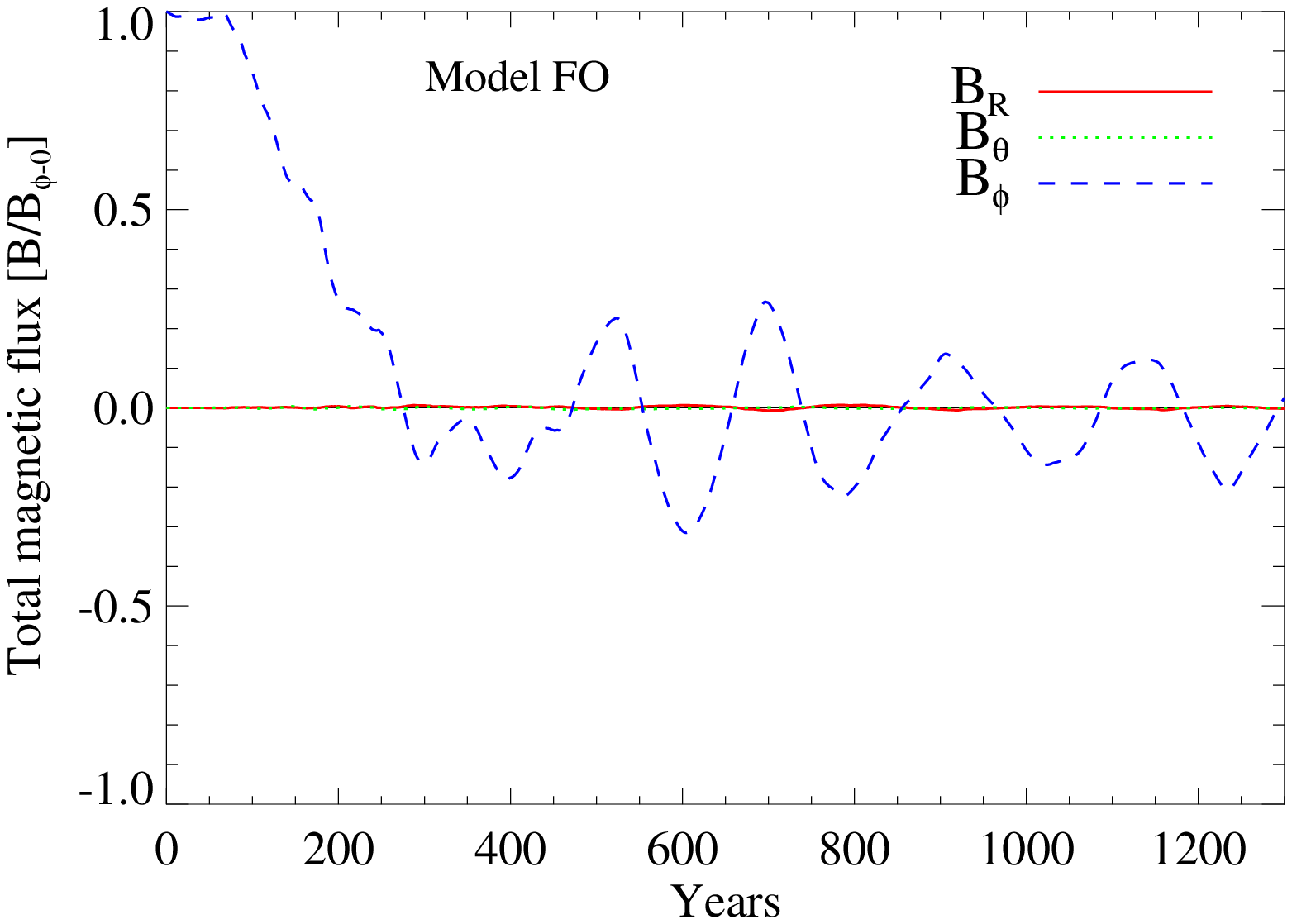,scale=0.55}
\psfig{figure=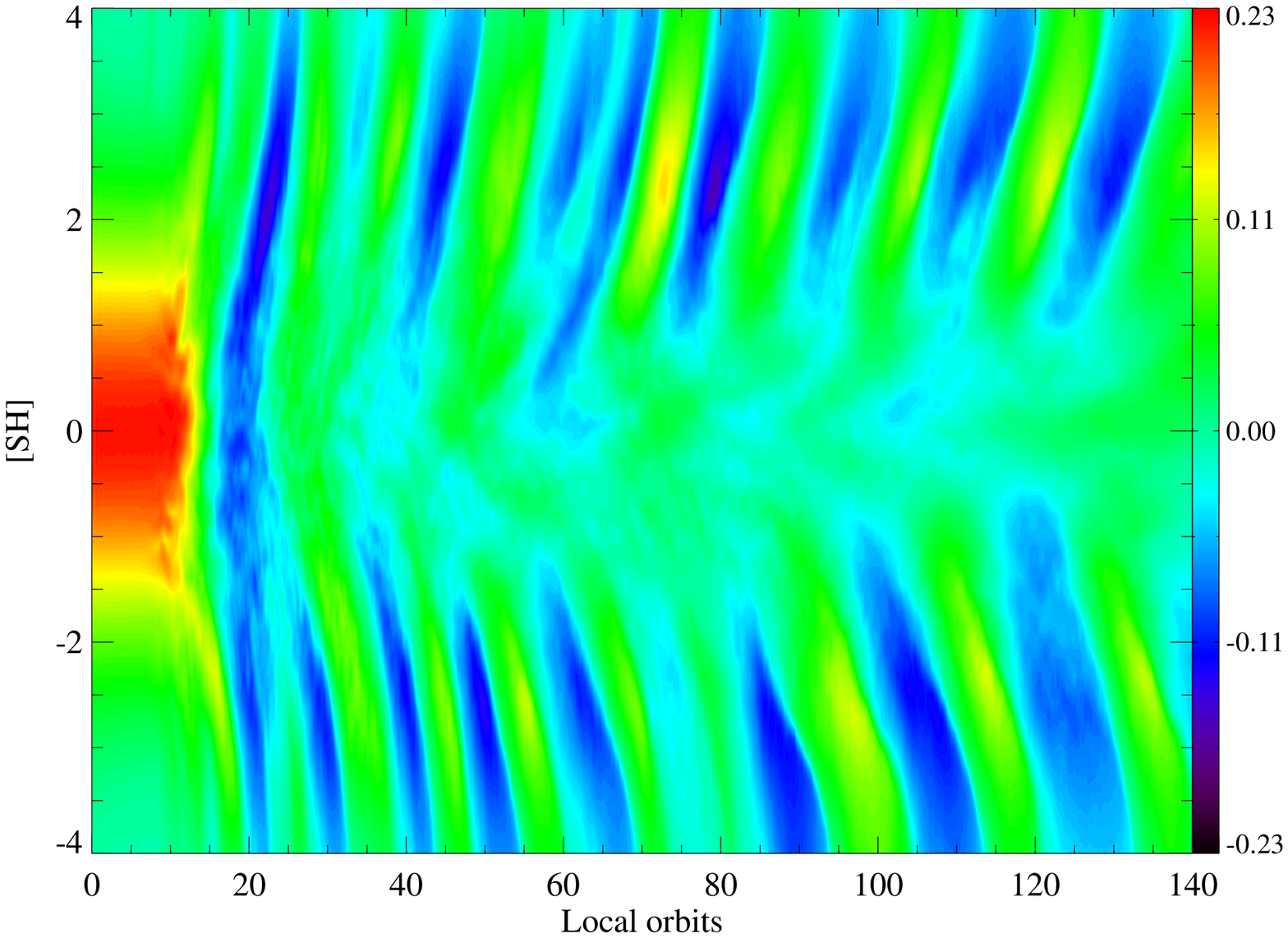,scale=0.46}
\end{minipage}
\hspace{4.0cm}
\begin{minipage}{5cm}
\psfig{figure=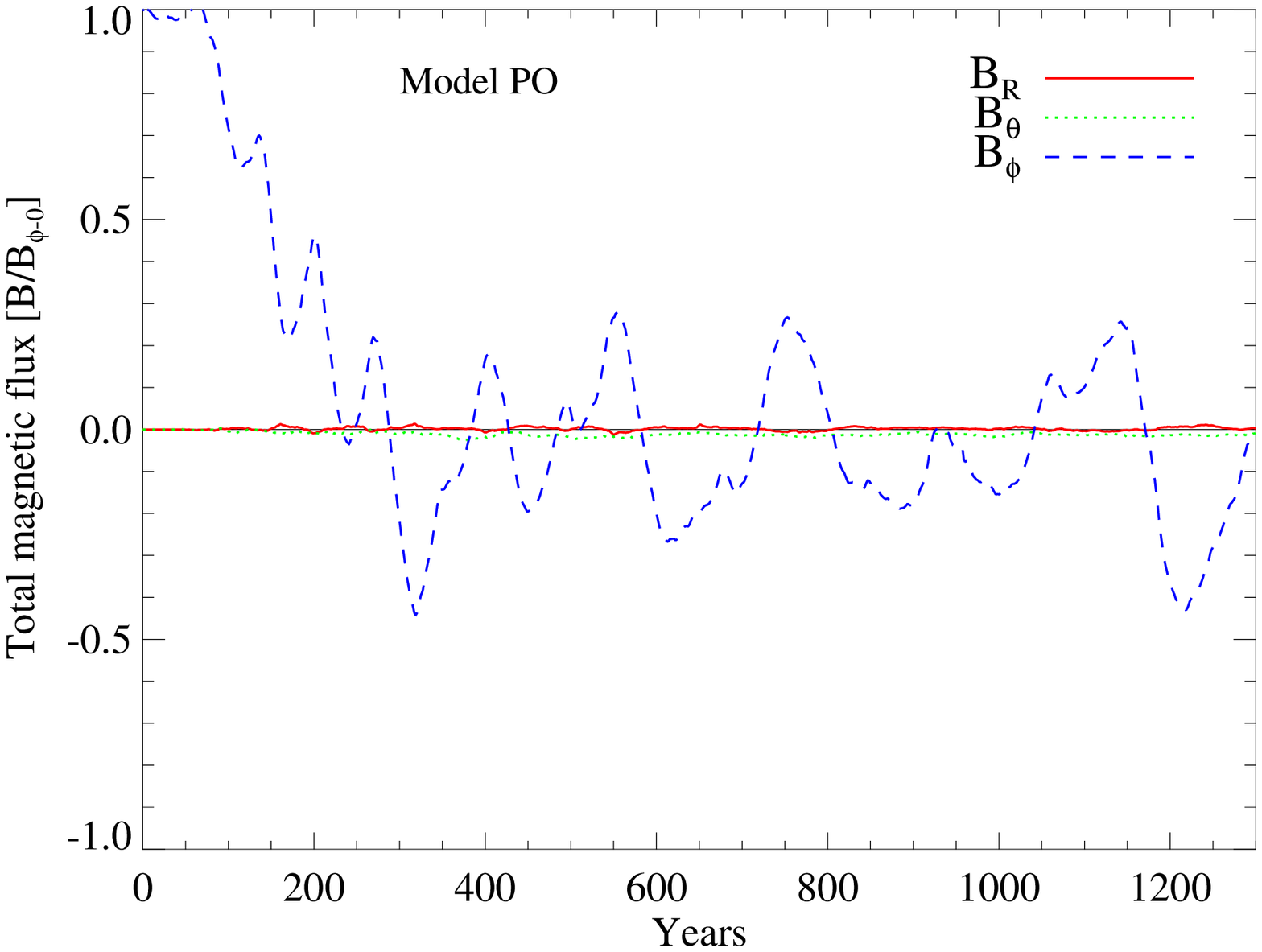,scale=0.46}
\psfig{figure=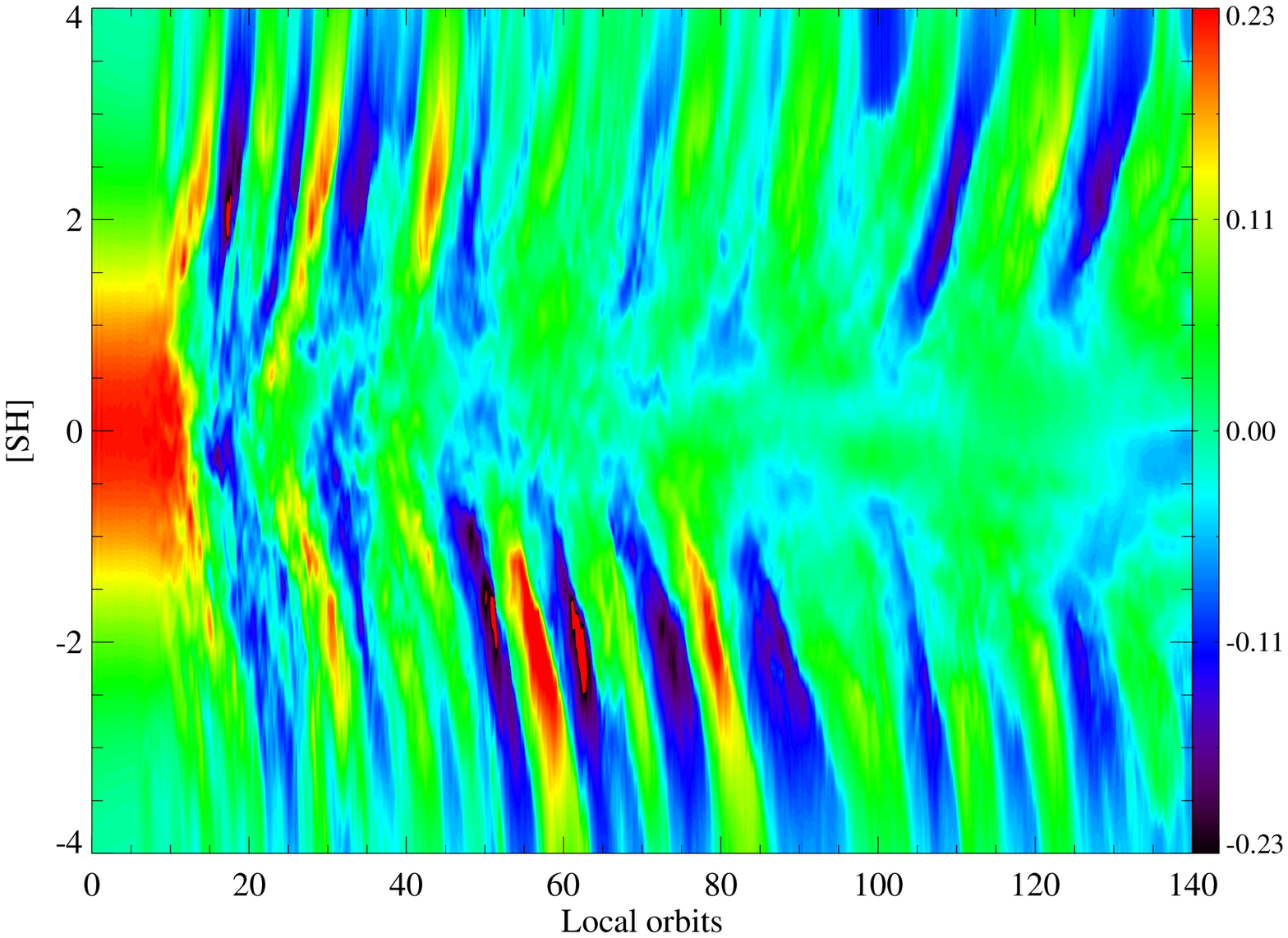,scale=0.46}
\end{minipage}
\label{mag_spec2}
\caption{
Top left: Total magnetic flux evolution integrated over the entire computational 
domain (without the buffer zones) normalized over initial flux $B_\phi$
in the $2\pi$ run FO and $\pi/4$ run PO (top right).
Bottom: Contour plot of $B_{\phi}$ over disk height and time. The value is averaged over
azimuth and radius (4 AU to 5 AU). Local 
orbits are calculated at 4.5 AU. Bottom left: model FO. Bottom right: model
PO. The butterfly pattern becomes visible. The $\pi/4$ model shows irregular and stronger amplitudes.}
\end{figure}

\section{Discussion}
For a number of aspects, our Godunov method confirms results previously
obtained with a finite difference method as presented by \citet{fro06}.
\begin{itemize}
\item A minimum amount of grid cells per scale height, which is about 25 grid
cells, is needed to sustain turbulence, which was to be expected as both methods 
present similar numerical behavior \citep{flo10}. 
Otherwise the turbulent magnetic energy slowly decays in the nonlinear MRI
evolution \citep{fro06}. 
Our highest resolution model BO was able to sustain a constant level of
turbulent stress for more than 400 inner orbits.

\item The toroidal magnetic net flux is quickly lost via an 
      open vertical boundary. Then, there is a oscillating zero-net
flux field present in the disk.

\item The disks show a two layer structure of turbulence.
\end{itemize}

\subsection*{$\alpha$-stress evolution}
We obtained a steady state $\alpha$ value of about $5\cdot10^{-3}$, which
is comparable to the results obtained in \citet{fro06}.
The time averaged radial profile of $\alpha$ follows $\sqrt{r}$.
This profile can be explained by the choice of our
radial pressure (density and temperature) profile, in combination
with the resulting magnetic field profile which is
force free, $|B'_\phi| \propto r^{-1}$.
For this magnetic field profile, the net radial magnetic force
vanishes. 
Any quasi steady state of disk turbulence must display this profile,
otherwise large scale radial readjustments in the density profile would occur.
Both $B'_\phi$ and $B'_r$ determine the Maxwell stress $B'_\phi B'_r $ to be $\sim 1/r^2$.
For the chosen $ \partial \ln{P} / \partial \ln {R} = -2.5$ this results in  $\alpha \sim B'_\phi
B'_r/P \sim \sqrt{r}$.\\
Of course, this profile is only valid for well-ionized accretion disk
regions.
The radial $\alpha$ profile in protoplanetary disks remains an open
question. 
The ionization rate, and possibly the MRI
activity, will be a function of radius and height \citep{sem04,dzy10}. Furthermore, the pressure scale height 
will vary with radius, and this also changes the MRI evolution.
Both effects will lead to different saturation levels for MRI turbulence
at different radii and thus also to different $\alpha$ values. 
\subsection*{Magnetic energy convergence}

\citet{dav10}, \citet{shi10} and \citet{fla10} show in local box simulations 
that the large scale magnetic energy converges for a resolution
between 32 and 64 grid cells per pressure scale height (\citet{dav10}, Fig. 3). 
Otherwise, the large scale magnetic energy decreases with increasing
resolution.
At the current state we could only handle 25 cells per scale height in our global
simulations. We expect also a large scale 
convergence of magnetic energy with 1.5(2.5) higher resolution.
Future calculations shall complete this point, but will be five times more computationally
expensive.\\
We can already conclude that our global magnetic energy spectra as well as the effect of increasing 
resolution (Fig, 12, bottom left, FO to BO) look very similar
to the results presented in local box simulations. 
The magnetic energy power spectra reveals that most of
the magnetic energy is deposited in the small turbulent scales.
For the model with a restricted azimuthal domain of $\pi/4$,
the largest energy scale is always the domain size.
%
\subsection*{Turbulent velocity}
Recently observed turbulent velocities in TW Hya and HD 163296
\citep{hug10} fit nicely to our computed Mach numbers of 0.1 and 0.4 in the midplane and corona of the
disks.\\
The kinetic energy spectra as well as the velocity spectra along
azimuth ($k_\phi$ space) show a peak between $\rm m=5$ and $6$ due to radial shear
waves.
Latest results in local box MRI simulations presented
a $k^{-3/2}$ slope for the kinetic energy power spectra \citep{fro10}.
Similar as in \citet{fro10}, our power-law fit 
applies only for a small range in k space.
We do not find a Kolmogorov type slope of $k^{-5/3}$ (Fig. 6).
Further studies are needed, including the $k_r$ and $k_{\theta}$.\\
A Kolmogorov scaling was predicted for magnetic ISM turbulence by
\citet{gol95} and recently confirmed in numerical simulations by \citet{ber10}.
However, it only applies for the inertial range of incompressible
isotropic turbulence. The driving of the turbulence via MRI,
the anisotropy of the turbulent eddies, the geometry and rotation of the disk 
and the compressibility of the gas
make it difficult to argue for a Kolmogorov scaling. 
We expected a spectrum to be more or less deviating from this simple law.

\subsection*{Two-phase disk structure}
We observe that the accretion disks establish a two-phase structure:\\
- The midplane region between $\pm$ two scale heights shows a pretty
constant turbulent RMS velocity of about $10\%$ of the local sound speed
 independent of radius or height.
Part of the RMS velocity occurs due to global shear waves which have radial
peak velocities of up to $30\%$ of the local sound speed. 
The amplitude of the azimuthal fluctuations in the magnetic
field is also independent of height ($\pm$ two scale heights around the midplane) but develops a 
$1/r$ profile in radius.
The midplane region shows a broad distribution of plasma beta values, 
$\beta = \frac{2P}{B^2}$, with a mean value of about 500 and a full width at
half maximum of two order of magnitude.\\
- In the coronal region, more than two scale heights above the midplane,
 the mean turbulent velocity reaches a Mach number of 0.5 with supersonic peaks up to 1.5. 
The mean magnetic fields decrease in this region with height.
The disk corona shows a narrower distribution of the plasma beta values with most values between 1
and 10. Here the magnetic fields are buoyant, gas and fields are
expelled from the disc. 
Relative high plasma beta ($\beta > 1$) in the corona have been
reported in \citet{fro06} for global models of AMRI with open boundary. The
magnetic flux escapes through the vertical boundary with a remaining zero-net flux
in the computational domain. This leads to the weakly magnetized corona (below equipartition).\\

\subsection*{Vertical outflow}
Our models show a MRI driven vertical outflow.
Above 2 scale heights, the gas flow is directed vertically and radially 
outward, Fig. 3. The outflow velocity of the gas (measured at the vertical
boundary) is still subsonic.
The disk evaporation time was determined at 5 AU to 2000 local orbits.
The launching region is located between 1.6 and 2 scale heights. 
This results matches values obtained in local box simulations \citep{suz09,suz10} with a
vertical net-flux field.\\
However, we are aware that a detailed study of the vertical outflows
requires 
much broader vertical extended
simulations to confirm that the gas is evacuated from the disk and not
returning.
These simulations should then include the sonic point or even the Alfv\'enic point to give
further insight into disk-wind and disk-jet interacting regions.

\subsection*{Meridional flows}
Our present work shows that the meridional
outflow at the midplane is only present in HD simulations, e.g., in
viscous simulations with an $\alpha$ value assumed to be constant in time and space.
For our MHD models, we find time variations of the orbital frequency of
around 50 local orbits, which are not present in the viscous disk models
and which prevent a steady radial outflow.
A similar result, the absence of a meridional flow in global MHD simulations 
was recently found by \citet{fro11}.\\
However, we confirm the more general picture of a viscous disk and 
show that viscous disk models with a radial viscosity
profile can reproduce successfully the radial mass flow rate in global MRI turbulent stratified disks.
Clearly, the vertical mass flow cannot be described with such an HD
model.\\

\subsection*{Mean field evolution}
The azimuthal MRI is self-sustaining in our zero net flux simulations with
open boundaries.
The fact that the total flux oscillates around zero could be due
to the generation of a mean poloidal magnetic field by a turbulent toroidal
field. 
We observe also an antisymmetric distribution of the mean magnetic fields in the upper and
lower hemisphere which could be an indication for the action of an MHD dynamo in our global
simulations.\\
The existence of an $\alpha$-$\Omega$ MHD dynamo and its role for accretion
disks was investigated by \citet{bra95}, \citet{zie00}, \citet{arl01},
\citet{bra07}, \citet{les08} and \citet{bla10}.
The temporal oscillations of the mean azimuthal field, plotted 
over height and time (Fig. 13), generates a butterfly pattern. 
Latest results connect the butterfly pattern with 
a dynamo mechanisms \citep{gre10}. 
We present also a butterfly pattern with a period of 10 local orbits,
independent of the azimuthal extent.
Additionally, the butterfly structure is reflected in the temporal
spatial fluctuations of the mean turbulent stresses with double period. 
A change of sign of the mean azimuthal field occurs every five local orbits, at the same time the 
$\alpha$-stresses show a minimum (Fig. 1, bottom right).\\
The magnetic energy as well as the mean field evolution have shown
that the $\pi/4$ model does not capture the correct properties 
of the larger scale simulations. 
\citet{haw00} also studied full $2\pi$ and restricted $\pi/2$ models of
accretion tori. However, a detailed study of the impact of different 
azimuthal domain extents is still needed and will be covered in future work.

\section{Summary}
We have performed full $2\pi$ 3D stratified global MHD simulations of 
accretion disks with the Godunov code PLUTO.
Our chosen disk parameter represent well-ionized proto-planetary disk
regions. We obtain a quasi steady state zero-net flux MRI turbulence 
after around 250 inner orbits.

\begin{itemize}

\item The second order Godunov scheme PLUTO including the HLLD Riemann
solver presents a similar nonlinear MRI evolution as finite difference
schemes. There is also a need of about 25 grid cells per pressure
scale height to reach a self-sustaining MRI turbulence in global zero net flux
azimuthal MRI simulations.

\item We observe a total $\alpha$ parameter of about $5\cdot10^{-3}$,
which remains constant for at least 400 inner orbits and scales with $\sqrt{r}$ for our used pressure and
density profile.

\item The turbulent magnetic fields show a $1/r$ profile in radius, mainly
visible in the dominating toroidal magnetic field. This configuration is force free
in the sense that there exist no large scale net force on the gas. This
profile determines the slope of the $\alpha$ parameter.

\item The magnetic energy spectra is similar as in local box simulations.
Most of the magnetic energy is placed in the smallest resolved turbulent
scale. 

\item The kinetic energy spectra as well as the velocity spectra peak for an
azimuthal wavenumber between $\rm m=3$ and $5$ due to shear waves, driving the
radial velocity up to a Mach number of 0.3. We do not find a Kolmogorov type
scaling in the $k_\phi$ space.

\item The model with an azimuthal extent of only $\pi/4$ has most of the energy at the domain
size and does not show the same mean field evolution. 

\item We observe a butterfly pattern with then local orbits independent of
the azimuthal extent. The butterfly period becomes also visible in the
Maxwell stress with double period. The mean magnetic fields are
antisymmetric for the two hemispheres.

\item At the midplane ($\pm 2 $ disk scale heights), 
our turbulent RMS velocity presents a constant Mach number of 0.1
independent on radius. At the corona ($> 2$ disk scale heights), the
turbulent velocity increases up to a Mach number of 0.5 at 4 scale heights.

\item The turbulent magnetic fields at the midplane present a broad plasma
beta distribution with a mean of about 500 $\pm$ one order of magnitude.
In the corona the plasma beta is between unity and ten.

\item The turbulent and the mean velocities are pointing vertically and
radially outward in the disk corona ($> 2$ disk scale heights). We observe a
steady vertical outflow for the open boundary models, dominating the radial
accretion flow.

\item We do not see a meridional flow pointing radially outward at the
midplane in our MHD models. However, we reproduce our total radial mass flow
in 2D viscous disk simulations with a radial dependent $\alpha$-viscosity.

\end{itemize}

\section{Outlook}

This paper presents a huge data set of about 10 TBytes.
This means we will continue to analyze the data for different goals. 
One study will deal with a closer investigation of dynamo properties in
our global disk models. Another one will analyze the turbulent spectra in a
better way to derive correlation times for the turbulence.
We will also fill the parameter space with $\pi$ and $\pi/2$ models to
identify whether a subsection will be sufficient. Higher resolution is
envisioned to reach resolution per scale height comparable to recent stratified local
box simulations. Finally, our global MHD model will be the work horse for our
future investigations of planet formation processes in circumstellar disks,
like collisions of boulders, planetesimal formation and planet migration.\\
In future runs, we also plan to use non-ideal MHD to include more realistic
magnetic Prandtl numbers and magnetic Reynolds numbers to understand the occurrence and
saturation level of the turbulence. Improving the thermodynamics is also
a must in future work, dealing with the proper ionization of the disk, like
capturing $p\Delta V$ terms or magnetic dissipation as heat input.

\acknowledgments
We thank Andrea Mignone for very useful discussions about the numerical setup.
We thank Sebastien Fromang for the helpful comments on the global models and 
on the manuscript.
We thank Alexei Kritsuk for the discussion about turbulent spectra.
We also thank Willy Kley for the comments on the viscous model.
We thank Frederic A. Rasio and an anonymous referee for the fast and very professional processing of
this work.
H. Klahr, N. Dzyurkevich and M. Flock have been supported in part by the
Deutsche Forschungsgemeinschaft DFG through grant DFG Forschergruppe 759
"The Formation of Planets. Neal Turner was supported by a
NASA Solar Systems Origins grant through the Jet Propulsion
Laboratory, California Institute of Technology, and by an Alexander
von Humboldt Foundation Fellowship for Experienced Researchers.
The Critical First Growth Phase". Parallel
computations have been performed on the PIA cluster of the MaxPlanck
Institute for Astronomy Heidelberg as well as the GENIUS Blue Gene/P cluster
both located at the computing center of the MaxPlanck Society in Garching.
\bibliographystyle{apj}
\bibliography{floc}
\end{document}